\documentclass[12pt,letterpaper]{article}
\usepackage[a4paper, total={7in, 10in}]{geometry}

\usepackage{graphicx}
\usepackage{helvet}
\usepackage{authblk}
\usepackage[colorlinks, linkcolor=blue]{hyperref}
\usepackage[tbtags]{amsmath}
\usepackage{amssymb}
\usepackage[super,comma,sort&compress]
   {natbib}

\usepackage[autostyle]{csquotes}
\usepackage[utf8]{inputenc}
\usepackage{subcaption}
\usepackage[newcommands]{ragged2e}
\usepackage{comment}
\usepackage{xcolor}

\usepackage{amssymb}
\usepackage{gensymb}
\usepackage{float}
\usepackage{amsmath}
\usepackage{braket}
\usepackage{tabularx,graphicx}
\usepackage{epstopdf}
\usepackage{latexsym}
\usepackage{color, colortbl}
\usepackage{psfrag}
\usepackage{bbm}
\usepackage{bm}
\usepackage{titlesec}
\usepackage{dsfont}
\usepackage{feynmp}
\usepackage{slashed}
\usepackage{xr}

\usepackage{multirow}
\usepackage[normalem]{ulem}

\def \vp {\varphi}

\def \beq {\begin{eqnarray}}
\def \eeq {\end{eqnarray}}
\def \tn {\textnormal}

\def \la{\langle}
\def \ra{\rangle}

\def \nn {\nonumber}

\def \h {\mathcal{H}}
\newcommand{\calH}{\h}
\newcommand{\hn}{\calH^0_{\widehat{n}}}
\newcommand{\hf}{{\widehat{\varphi}}}
\definecolor{applegreen}{rgb}{0.5, 0.8, 0}
\newcommand{\rd}{{\rm d}}

\makeatletter
\renewcommand{\maketitle}{\bgroup\setlength{\parindent}{0pt}
\begin{flushleft}
  \textbf{\@title}

  \@author
\end{flushleft}\egroup}
\makeatother

\title{Arresting Quantum Chaos Dynamically in Transmon Arrays}
\author[1,2,3]{Rohit Mukherjee}
\author[1,3]{Haoyu Guo}
\author[1]{Keiran Lewellen}
\author[1,4,*]{Debanjan Chowdhury}

\affil[1]{Department of Physics, Cornell University, Ithaca NY 14853.}
\affil[2]{Department of Physics, Indian Institute of Technology, Kanpur 208016, India}
\affil[3]{These authors contributed equally to this work}

\affil[4]{Lead contact}
\affil[*]{Correspondence: debanjanchowdhury@cornell.edu}

\makeatletter
\newcommand{\customlabel}[2]{%
   \protected@write \@auxout {}{\string \newlabel {#1}{{#2}{\thepage}{#2}{#1}{}} }%
   \hypertarget{#1}{}
}
\makeatother

\begin{document}

\maketitle

\section*{Summary}
Ergodic quantum many-body systems evolving under unitary time dynamics typically lose memory of their initial state via information scrambling. Here we consider a paradigmatic translationally invariant many-body Hamiltonian of interacting bosons --- a Josephson junction array in the transmon regime --- in the presence of a strong Floquet drive. Generically, such a time-dependent drive is expected to heat the system to an effectively infinite temperature, featureless state in the late-time limit. However, using numerical exact-diagonalization we find evidence of special ratios of the drive amplitude and frequency where the system develops {\it emergent} conservation laws, and {\it approximate} integrability. Remarkably, at these same set of points, the Lyapunov exponent associated with the semi-classical dynamics for the coupled many-body equations of motion drops by{  at least an} order of magnitude, arresting the growth of chaos. We supplement our numerical results with an analytical Floquet-Magnus expansion that includes higher-order corrections, and capture the slow dynamics that controls decay away from exact freezing.

\section*{Keywords}
Floquet many-body dynamics, Quantum thermalization, Quantum Chaos, Transmon devices, Floquet-Magnus expansion, Semiclassical dynamics

\section*{Introduction}

The thermalization dynamics of closed quantum many-body systems has been a major focus in recent decades \cite{Deutsch,srednicki,Tasaki,rigol}. In non-integrable systems, the eigenstate thermalization hypothesis (ETH) is central to the approach to equilibrium, where thermalization proceeds from local relaxation to scrambling, often driven by quantum chaos. However, thermalization can be interrupted in cases of strong disorder, leading to many-body localization (MBL) \cite{basko2006metal,gornyi,Huserev,Altmanrev,RMPMBL,resubref5}, or by the presence of quantum ``scar" states in constrained systems \cite{Serbyn2021,Scars_AR,DM6,resubref1,resubref2,PhysRevB.106.035123}. One of the appeals of arresting the onset of thermalization in many-body systems is their ability to preserve the memory associated with an initial state up to late times. However, the stability of MBL as a distinct phase of matter in the thermodynamic limit remains unclear \cite{roeck,PhysRevB.99.134305,vidmar,avalanches,sels}, and the fraction of scar states in the many-body spectrum of many-body Hamiltonians is vanishingly small.

A special class of thermalization dynamics arises in the context of driven many-body systems, such as time-periodic Floquet Hamiltonians. While generically such driven Hamiltonians thermalize to a completely featureless state with effectively ``infinite-temperature'' at asymptotically late times~\cite{PhysRevE.90.012110,PhysRevX.4.041048}, under special circumstances, the heating can be slow with a long regime of a pre-thermal phase~\cite{dimaprethermal1,dimaprethermal2,PhysRevB.98.104303,khemaniprethermal}.

In an important example of such phenomena, recent studies on strongly driven spin models with bounded local Hilbert spaces show evidence of dynamical ``freezing" at a sequence of special ratios of the drive amplitude and drive frequency, leading to approximate emergent conservation laws \cite{PhysRevB.82.172402,Haldar_2022,freezing1,freezing2,freezing3,freezing4,KS1,KS2,KS3,KS4,KS5,KS7,sreemayee,mukherjee2024instanton}. Interestingly, the location of the above freezing-points coincide with the points at which the leading contribution in a Floquet-Magnus expansion vanish. However, there are a number of questions that remain. This includes (i) the role of the local Hilbert space dimension, including an unbounded spectrum, on freezing, (ii) the structure of the higher-order corrections, which controls both the exact location of the freezing point and the residual slow relaxation at  freezing, and (iii) the nature of chaos, or lack thereof, at and near freezing. In an accompanying paper \cite{SYK_Freezing}, some of us have addressed a subset of these questions definitively in an exactly solvable zero-dimensional model of chaotic quantum dots. Furthermore, we also identified a general criterion for freezing, which will be tested in the present manuscript.

In this manuscript, we address many of these basic questions regarding dynamical freezing using a multi-pronged approach. We focus on a paradigmatic model of enormous contemporary interest, namely an array of Josephson junctions (Fig.~\ref{fig:Schemetic}), which forms the fundamental building block of superconducting ``transmon" based quantum computing ~\cite{RevModPhys.93.025005,PhysRevA.76.042319}. A single transmon setup consists of a Josephson junction and a shunting capacitance; the array is constructed out of capacitively coupled transmons. The transmon eigenspectrum is unbounded, but a few of the low lying eigenstates can be chosen to store quantum information. Recent work has indicated that such transmon arrays shows strong signatures of chaos~\cite{borner2023classical,berke2022transmon,resubref3,resubref4,resubref6}, which can become a future impediment towards scaling up the arrays to tens of thousands of qubits. As the transmons are excited to higher energies, the nonlinearities become more pronounced and trigger a transition from integrable to chaotic motion. In this work we propose that dynamical freezing can be used effectively for suppressing chaos in a large array of transmons, and perhaps other related qubit architectures in the future.

Presently, chaos is managed by either tunable couplers, as in Google’s Sycamore device \cite{arute2019quantum,Goggle}, or frequency detuning, as in IBM chips \cite{8936946}. While the first approach stabilizes the system, the additional elements in the circuit add architectural complexity and may introduce new channels of decoherence. In the second approach, the intentional frequency detuning (``disorder") effectively forces the system to become many-body localized, and thereby integrable. Whether this is a viable approach with increasing system volume is presently unclear. Regardless of the above protocol, finding new ways to dynamically protect the information stored in individual qubits in the NISQ era devices \cite{preskill} is a desirable but challenging goal~\cite{Gill7,nguyen2024programmable}.

\section*{Results}
\subsection*{Coupled Transmons}
\label{transmon}

The model of interest to us is an array of coupled transmons~\cite{berke2022transmon} in the presence of an external time-periodic drive. The total Hamiltonian is composed of three different parts,
\begin{subequations}
\beq\label{Eq:totaHam}
H(t) &=& \h^0_{\widehat{n}}+\h^0_{\widehat{\varphi}}+f(t)\h^{\rm{drive}}_{\widehat{n}},\\
\label{Hamiltonian1}
    \h^0_{\widehat{n}} &=& 4E_{c}\sum_{i}\widehat{n}_{i}^{2} + V\sum_{\langle i,j\rangle}\widehat{n}_{i}\widehat{n}_{j}, \\ \label{Hamiltonian2}
    \h^0_{\widehat{\varphi}} &=& -\sum_{i}E_{J_{i}}\cos{(\widehat{\vp}_{i})}\,,\\
    \h^{\rm{drive}}_{\widehat{n}} &=& \sum_i \widehat{n}_i,
\eeq
\end{subequations}
where $\widehat{n}_{i}$ is the Cooper pair number operator, $\widehat{\varphi}_{i}$  is the superconducting phase at site $i$ (which is compact and takes value in the range ($-\pi,\pi]$) ,
$E_{c}$ is the capacitive charging energy, $E_{J_{i}}$ represent the Josephson energies of the individual transmons (in principle, drawn from some narrow normal distribution), and $V$ is the coupling between nearest-neighbor transmons; see Fig.~\ref{fig:Schemetic}. Physically, $E_{J}$ corresponds to the ability of Cooper pairs to tunnel across the junction. The canonically conjugate operators, $\widehat{n},~\widehat{\vp}$ obey the usual commutation relation $[\widehat{\varphi}_{i},\widehat{n}_{j}]=i\delta_{ij}$, and can be expressed as~\cite{girvin2014circuit},
\beq
    \widehat{\vp}_{i}=\Big(\dfrac{2E_{c}}{E_{J_{i}}}\Big)^{1/4} [a_{i}^{\dagger}+a_{i}],~
    \widehat{n}_i=\dfrac{i}{2}\Big(\dfrac{E_{J_{i}}}{2E_{c}}\Big)^{1/4}[a_{i}^{\dagger}-a_{i}],
\eeq
where $a_i^\dagger,~a_i$ represent bosonic creation and annihilation operators, respectively. Note that the spectrum and the associated local Hilbert space for the static Hamiltonian is unbounded from above, which {\it a priori} makes the fate of any freezing type phenomenology far from obvious. In particular, the nature of Fock-space (de-)localization, quantum interference, and approach to equilibration via chaos, can be distinctly different in the transmon model compared to spin-models. Interestingly, a Floquet drive-induced many-body localization has been analyzed in a one-dimensional weakly disordered bosonic Hamiltonian \cite{Lindner1}, where the drive effectively suppresses the static hopping amplitude \cite{eckardtrmp} and pushes the system towards the MBL phase. In contrast, this work focuses{  primarily on} translationally invariant Hamiltonians, where the static many-body system is fully ergodic and the emergent (approximate) integrability arises via a distinct mechanism.

For simplicity, we have ignored the offset charge(s) \cite{roth2021introduction}, $n_g(t)$, which are known to be a source of decoherence in charge qubits ~\cite{fluxonium1,PRXQuantum.2.030101}. In what follows and unless otherwise stated, we will assume a translationally invariant Hamiltonian with uniform Josephson energies, $E_J$.{   We will comment on the results of breaking this translational invariance in Sec. \ref{sec:IBM_simulations}.} We will restrict our attention to the usual transmon regime characterized by $E_{J}\gg E_{c}$ with an eye towards making connection to present day hardware, but the freezing phenomenology is not solely restricted to this hierarchy of energy scales.

Now we turn our attention to the main ingredient, the Floquet drive, $f(t)$ with a period $T$, such that $f(t) = f(t+T)$; equivalently the drive frequency, $\omega = 2\pi/T$. For the sake of concreteness, we will focus on two types of drives (square wave and cosine wave, respectively) with an amplitude $A$ in the remainder of this study:
\begin{subequations}
\beq
{\rm{Case~ I:}}~~f_{\rm{I}}(t) &=& A~{\rm{sgn}}[\sin(\omega t)]\label{drive1},\\
{\rm{Case~ II:}}~~f_{\rm{II}}(t) &=& A\cos(\omega t).\label{drive2}
\eeq
\end{subequations}
Our focus in the remainder of this manuscript will be on the quantum dynamics, chaos and freezing (or lack thereof) for the model introduced in Eq.~\eqref{Eq:totaHam} due to the drive in Eq.~\eqref{drive1}-\eqref{drive2} as a function of $A/\omega$. We will start by presenting results for the exact diagonalization in Sec.~\ref{results:ED} below. Given the unbounded local Hilbert-space, ${\cal{H}}_D$, we will analyze the eigenspectrum of $H(t)$ with a varying truncation scheme while expanding the $\cos(...)$ non-linearity in powers of $\widehat\vp$; see Supplemental Methods \ref{App:hilbert_dim} for details.

For the numerical results presented in the remainder of this manuscript, we will use the IBM parameters~\cite{IBM},
{  $E_{J}/h=12.58$ \ \text{GHz} , $E_{c}/h=330$ \ \text{MHz}, $V/h=50$ \ \text{MHz}. Here to clarify the units we have explicitly restored Planck's constant $h$, but throughout the remainder of the text we will work in units of $h=1$.} The transmon regime is defined by $E_{J}\gg E_{c}$, where the{  oscillator frequency for each transmon is approximately given by }{ 
\begin{equation}
h\nu_{i}=\sqrt{8E_{J}E_{c}}\label{eq:transmon_frequency}
\end{equation} and is of the order of $\nu\sim 5$ GHz.}

\subsection*{Exact Diagonalization}
\customlabel{results:ED}{(\textbf{Exact Diagonalization})}
We begin by analyzing the emergent conservation law tied to $\widehat{N}$ from the ED computations. In this regard, the quantity of principle interest is the expectation value,
\begin{equation}
    \la\widehat{N}\ra= \frac{1}{L} \sum_{i}^{L}\bra{\psi(t)}\widehat{n}_{i}\ket{\psi(t)},\label{Eq.N}
\end{equation}
 starting from a specific initial state $\ket{\psi(0)}$, which yields the associated time evolved state, $\ket{\psi(t)}$. Freezing is identified by the asymptotic late-time behavior of this quantity, which is analyzed as a function of the parameter $A/\omega$. Recall that without the drive, the Hamiltonian $H_{0}\equiv{\cal{H}}_{\widehat{n}}^0 + {\cal{H}}_{\widehat{\varphi}}^0$ is known to be ergodic and chaotic~\cite{borner2023classical}, and since $\widehat{N}$ is not a conserved operator its expectation value is expected to relax rapidly. We also compute the time averaged (over Floquet cycles) quantity,
\beq
{\cal{F}} = \overline{ \langle \widehat{N}(t) \rangle / \langle \widehat{N}(0) \rangle},\label{Eq.Fidelity}
\eeq
at and away from freezing,{  which serves as a proxy for measuring the emergent conservation of $\hat{N}$ and}  will help determine the shift of the freezing point due to higher order corrections.{  Here $\langle...\rangle$ represents the expectation value with respect to some chosen state and the ``$\overline{...}$" corresponds to a time average over many Floquet-cycles.}

{ In the main text, we focus on a one-dimensional system of 8 transmons with a local Hilbert space dimension of 3 for our exact-diagonalization computations. However, we have thoroughly examined the freezing phenomenon by varying both the system size and the onsite Hilbert space dimension in Supplemental Methods \ref{App:hilbert_dim}. Our findings indicate that the freezing point remains highly robust as a function of the varying system size and Hilbert space dimension.}

We initialize the computation in the time domain with different initial states, $\ket{\psi(0)}$, which are either eigenstates of $H(0)$ (which includes the drive term), or thermal superposition states constructed out of these, $\propto\sum_n e^{-\beta \epsilon_n}|n\rangle$. Unless otherwise stated, the default choice for the initial state is the ground state of $H(0)$. We notice that most eigenstates of $H(0)$, {\it except} the states in the middle of the spectrum, can be frozen at $A/\omega$ close to the prediction of the leading order Magnus  expansion. For many-body states in the middle of the spectrum, the expectation values for $\braket{\widehat{n}_i}$ become progressively smaller and the entanglement entropy saturates (see below) in a fashion that makes it numerically difficult to assess the extent of freezing.{  The ground state or the low temperature thermal superposition states of $H(0)$ have relatively large expectation values for $\braket{\widehat{n}_i}$, so by choosing to initialize in these state, we manage to avoid the numerical issues stated above.}

We also analyze the entanglement entropy (EE) of the system, at and away from freezing, for a variety of initial states. The half-chain EE is defined as,
\begin{equation}
    S_{\rm ent}=-\text{Tr}[\rho_{\frac{1}{2}}\log \rho_{\frac{1}{2}}],
\end{equation}
where $\rho_{\frac{1}{2}}$ is the density matrix of one half of the chain, computed by tracing out the other half. The entanglement dynamics between different parts of a system serves as a useful diagnostic for identifying the growth of quantum chaos~\cite{PhysRevE.70.016217}. A non-integrable Floquet system that thermalizes generically at late times, shows a saturating behavior for $S_{\rm ent}$. The saturation value depends on the  Hilbert space dimension, and is described by a volume law scaling~\cite{PhysRevLett.119.220603}. However, when the system becomes dynamically frozen, the system develops approximate integrability and leads to extremely slow dynamics. This behavior should be evident in the overall suppression and slow growth of $S_{\rm ent}$. Finally, we note that the exact nature of the residual dynamics in the frozen state can, in principle, be dependent on the choice of the initial state, which will be analyzed below.

In order to highlight the salient features of the freezing phenomenology that emerges under the square vs. cosine-wave drives, we discuss these separately below in the following two subsections.{  We will see that, while freezing occurs for both drives, it is quantitatively stronger in the case of the square-wave drive. This behavior aligns well with what our Magnus-expansion based theory predicts and stems from the unique form of the Fourier coefficients of the square-wave drive. (See Supplemental Methods \ref{app:FM} for a detailed derivation.)}

\subsubsection*{Square-Wave Drive}

In Fig.~\ref{fig:ndtstrobo}(a),(b),(c), we plot the stroboscopic time evolution for the operator $\widehat{N}$ (defined in Eq.~\eqref{Eq.N}),  normalized by its initial value. We start either with the ground state of the total Hamiltonian at time $t=0$, or a thermal superposition of the eigenstates of $H(0)$, defined as $\ket{\Psi(0)}=\sum_{n} \exp{(-\beta \epsilon_{n})}\ket{\psi_{n}}$. When the amplitude of the drive is away from freezing, the system loses memory of the initial state rapidly. On the other hand, at the freezing point, there is an approximate emergent conservation law associated with $\widehat{N}$, and the system retains its memory up to late times. However, we emphasize that this emergent conservation is only approximate, and in Sec.~\ref{results:FM} we calculate the corrections to freezing within a Magnus expansion, which are small in $\mathcal{O}(1/ \omega)$. Moreover, these corrections can also result in a shift of the location for freezing{  which is more pronounced in the cosine wave driving, that we will revisit using the Magnus expansion.} Interestingly, the corrections can depend on the choice of initial states, which therefore also affects the robustness of the associated freezing phenomenology.

In Fig.~\ref{fig:ndtstrobo}(d),(e),(f) we show the stroboscopic time evolution of the growth of half chain many-body quantum entanglement entropy ($S_{\rm ent}$) for different{  initial} states. Away from freezing,  $S_{\rm ent}$ grows rapidly and saturates to its equilibrated value. At freezing, the system becomes approximately integrable, and the growth of EE becomes slow while its saturation value depends on the Hilbert space dimension.{  In the inset of Fig.~\ref{fig:ndtstrobo}(d),(e),(f) we plot the relative change in entropy. With increasing temperature, which incorporates a superposition over a larger number of states from the many-body spectrum, the relative change increases at freezing but remains much smaller than away from freezing. Additionally, for various initial states away from freezing, the saturation value remains nearly the same, primarily dictated by the total Hilbert space dimension. }

\subsubsection*{Cosine-Wave Drive}

{ 
For the cosine drive, let us first present results for the emergent conservation of $N$ in Fig.~\ref{fig:cosonefidelity}, where we plot  $\mathcal{F}$ (defined in Eq.~\eqref{Eq.Fidelity}) for different choice of initial states. We find several salient features, which we now discuss one by one. First, compared to the square-wave drive case, we observe a sizeable renormalization of the location of the freezing point (marked by black arrows in Fig.~\ref{fig:cosonefidelity}(a)) compared to the prediction of the zeroth-order Magnus expansion (marked by dashed lines in Fig.~\ref{fig:cosonefidelity}(a)). In Sec.~\ref{results:FM}, we will show that such renormalizations of the freezing point are controlled by the corrections in the Magnus expansion, which also depends on the choice of initial state.

Second, the states in the middle of the spectrum are harder to freeze (see e.g. the less pronounced peak for the pink curve in Fig.~\ref{fig:cosonefidelity}(a)). There are two possible explanations for the reduced extent of freezing: (i) These initial states have a higher entanglement entropy (as shown later in Fig.~\ref{fig:cosinendt}) and hence longer range of entanglement. Therefore, the state is more susceptible to higher order corrections in the Magnus expansion, which consists of longer operator strings. (ii) These states have a smaller value of $\braket{\hat{N}}$, and is therefore more susceptible to fluctuations of $\hat{N}$.

Third, apart from the freezing peaks predicted by the leading-order Magnus expansion, we also observed secondary peaks that are located in between the freezing points. These peaks can also be explained within the Magnus framework. At these secondary locations, half of the non-integrable terms in the Hamiltonian are effectively switched off. A more detailed description of the scenario is given in  Supplemental Methods \ref{App:D}. }

In Fig.~\ref{fig:cosinendt}(a),(b),(c), we revisit the stroboscopic time evolution for $\widehat{N}(t)$ at the numerically determined renormalized freezing points (Fig.~\ref{fig:cosonefidelity}(a)). For $\ket{\psi(0)}$, we choose both the ground state and an excited eigenstate of $H(0)${  (as shown by the colored arrows in Fig.~\ref{fig:cosonefidelity}(b))};  $\widehat{N}$ becomes approximately conserved at the renormalized freezing points. The slow time decay associated with the emergent conservation in Fig.~\ref{fig:cosinendt}(c) is controlled by the higher order corrections in the Magnus expansion.

Finally, in Fig.~\ref{fig:cosinendt}(d),(e),(f) we show the stroboscopic evolution of the half-chain entanglement entropy. At the renormalized freezing points (indicated by red curves), there is a noticeable slow increase in the entanglement entropy compared to the square-wave drive. The difference stems primarily from the structure of higher-order corrections in a Magnus expansion; while the first-order correction for the square-wave drive is zero, it is non-zero for the oscillatory drive. However, away from freezing (represented by the blue curves), the system rapidly thermalizes, and the difference from the frozen curve is striking.{  The noticeable growth of entanglement entropy in this case even at the freezing point highlights the influence of interactions as was pointed out in  Ref.~\cite{Haldar_2022}.}

{  The exact diagonalization computations have been performed here for a finite system and Hilbert-space dimension, respectively, both of which introduce an exponential complexity. We have studied the systematic effects of both of these finite-size corrections on the freezing phenomenology in Supplemental Methods\ref{App:hilbert_dim}. Additionally, incorporating a time evolution axis makes the numerics challenging. For larger systems, where exact diagonalization becomes unfeasible, alternative methods like tensor network techniques, such as the infinite time-evolving block decimation (iTEBD) algorithm, remain an exciting avenue for future studies.}

\subsection*{Semi-Classical Limit}
\customlabel{results:SC}{(\textbf{Semi-Classical Limit})}

The transmon system we consider is amenable to a semi-classical treatment where we take $\hat\phi$ and $\hat n$ as real variables rather than quantum operators; see methods section \ref{sec:SClimit} for details. In this section, we present results of this semi-classical analysis. The coupled semi-classical equations of motion in Eq.~\eqref{SC1}-\eqref{SC2} for $\varphi_i$ and $n_i$ can be solved using a Tsitouras 5/4 Runge-Kutta method~\cite{tsitouras2011runge}, starting from a specific initial state. Our main goals in this section will be to analyze the chaotic properties associated with the resulting trajectories and observe their dependence on system size and varying amounts of disorder.

In order to pick a starting state that has a close correspondence with realistic ``computational states" in the quantum problem, we adapt the strategy highlighted in Ref.~\cite{borner2023classical}. First we calculate the eigenenergies, $E_{m}$, for a single static quantum transmon. The energies of the computational states are given by $E_0, ~E_1$ for the $|0\ra,~|1\ra$ state, respectively. We then initialize the transmons in the semiclassical limit as $0$ or $1$ by initializing their phase space coordinates as $(n,\varphi)$=$(0,\varphi_{m})$, where $-E_{J}\cos{\varphi_{m}}=E_{m}$ for $m=0,~1$.{  We have performed additional simulations that suggest that the semi-classical freezing phenomenology discussed below is only strengthened by initializing to non-zero values of $n$.} In this fashion we study the evolution with an initial semi-classical state $101010\cdots$. The{   maximum Lyapunov exponent} is computed using the ``H2 method" ~\cite{benettin1980lyapunov1,bennetin1980lyapunov2} and implemented in Julia ~\cite{datseris2018dynamicalsystems}.

{  \subsubsection*{Phenomenology of Semi-Classical Freezing}}
We begin first by discussing the results for the square-wave drive in Eq.~\eqref{drive1}. In Fig.~\ref{fig:classicalLE}(a) we plot the MLE as a function of the drive amplitude, $A$, for a fixed drive frequency, $\omega$. Remarkably, whenever $A/\omega$ approaches the special points in Eq.~\eqref{condn1},  $\lambda_{L}$ is suppressed by{   over an order of magnitude} compared to the value in the static case (shown by dashed line).{   To further highlight the correspondence between the quantum and classical results, we analyze freezing as a function of the drive frequency, $\omega$, in both the quantum and semi-classical cases. The Magnus expansion, which can account for many of our numerical results, is formally a large $\omega$ expansion, so we might expect that there exists a threshold for $\omega$ below which freezing no longer occurs. To see this, we evaluate $\langle \widehat{N}(t)\rangle$ (normalized by its initial value) from the ED data, and $\lambda_L$ calculated from the semi-classical equations for the same system size and parameters. We plot both as a function of $\omega/\nu$ where $\nu$ is the fundamental transmon frequencies defined in Eq.~\eqref{eq:transmon_frequency}. The results are shown in Fig.~\ref{fig:classicalLE}(b) at a fixed value of $A/\omega=2$, which corresponds to one of the freezing points. Both sets of results suggest that freezing requires the drive frequency to be above a threshold set roughly by the individual transmon frequency (single particle energy scale). We have also noticed that for a given value of $\omega$, freezing requires the drive amplitude to also be larger than a critical value. For a more in depth discussion of the $\omega$ dependence of semi-classical freezing, see Supplemental Methods \ref{App:semiclassical}.}

To help visualize how the trajectories for $\{\varphi,~n\}$ evolve with varying $A/\omega$ for the square-wave drive,  we generate phase-space portraits for two transmons (picked at random) from the total system consisting of $L=8$ transmons. When the system is away from freezing, and starting from an initial state $1010...$, the phase-space fills up uniformly over time and displays ergodic behavior as shown in Fig.~\ref{fig:phasespace8}(a). Specifically, $n$ displays strong fluctuations over time starting from its initial value. In contrast, at freezing the individual transmons exhibit strong deviations away from any ergodic behavior (Fig.~\ref{fig:phasespace8}(b)), leading to a highly constrained set of trajectories in phase-space showing minimal fluctuations of $n$.

Finally, turning to the cosine drive, Fig.~\ref{Lyapnuov_cosinedrive} shows a plot of $\lambda_L$ as a function of $A/\omega$ for an $L=8$ transmon array. Once again, we find that $\lambda_L$ drops significantly at freezing compared to the static Hamiltonian.{  However, the location of the freezing points is somewhat shifted from the predictions of the zeroth-order Magnus expansion. This was noted in the ED analysis as well, highlighting the significant role played by the higher order corrections in the case of the cosine drive.}\\

{\subsubsection*{Dependence on System Size and Disorder}}\customlabel{sec:IBM_simulations}{(\textbf{Dependence on System Size and Disorder})}

{  In this section, we analyze the extent of dynamical freezing within our semi-classical setup in systems of up to $433$ transmons arranged in the same geometries that IBM uses for its existing quantum devices. Clearly, this lies beyond the range of exact diagonalization, but given the excellent correspondence between the two sets of computations for small system-sizes, demonstrates the likely fate of freezing for larger systems. We then turn to the effect of disorder in the Josephson energy coupling on freezing.

The particular two-dimensional ``heavy-hexagon" geometries that IBM \cite{IBM} uses in its currently available quantum devices are shown in Fig.~\ref{fig:IBM_simulations}(a). For each of the Falcon, Hummingbird, Eagle, and Osprey devices, we initialize the semi-classical system in a checkerboard pattern of $1$'s and  $0$'s. In Fig.~\ref{fig:IBM_simulations}(b), we plot the MLE as a function of the drive amplitude $A$ at a fixed frequency $\omega$. The freezing phenomenology discussed above persists for all devices considered with $\lambda_L$ being significantly suppressed at freezing points. Furthermore, we find that with increasing system size, the growth of $\lambda_L$ is larger away from the freezing points.

So far we have worked exclusively with translationally invariant Hamiltonians. However, realistic transmon arrays have disordered values of $E_{J_i}$, which are either intentionally engineered or due to unintended imperfections. To study the effect of disorder on dynamical freezing in the semi-classical limit, we consider a one-dimensional array of length $8$ with $E_C = 330$ MHz, $V = 50$ MHz, and $E_{J_i} \sim \mathcal{N}(12.58, \sigma^2)$ GHz. Here, the values of $E_{J_i}$ are chosen from a normal distribution with standard deviation $\sigma$~\cite{IBMdisorder}, independently at each site. To ensure we stay in the transmon regime, we impose a cutoff of $4$ GHz as the minimum value of $E_{J_i}$. After driving the system with the square-wave drive at $\omega/(2\pi) = 15$ GHz and calculating $\lambda_L$, we average over $1000$ disorder realizations. In Fig. \ref{fig:Disorder_averaging} we plot $\lambda_L$ as a function of $\sigma$ for $A/\omega = 0, 2.17, 4.04$ -- respectively the undriven case and the numerically-calculated renormalized values for the first and second freezing points. We find that for small values of $\sigma$, the phenomenon of freezing persists. However, much larger values of $\sigma$ weaken freezing (particularly for the second freezing point) while simultaneously reducing chaos in the undriven case (the latter behavior being in alignment with previous results \cite{borner2023classical}). Nonetheless, we expect that in the asymptotic sense, for any given strength of disorder, a strong enough drive frequency can be used to reinstate the freezing phenomenology in principle. However, we leave a detailed study of this for future work.}\\

\subsection*{Floquet-Magnus Expansion}
\customlabel{results:FM}{(\textbf{Floquet-Magnus Expansion})}

In this section, we extend the Floquet-Magnus expansion for the transmon Hamiltonian beyond the leading order, as described in the methods section \ref{sec:FM}. In particular, we use the expansion to compute the renormalized freezing point. As will become clear later, it will be useful to consider the drive amplitude to vary across different sites. Thus we modify the drive term as,
\begin{equation}
\calH_{\widehat{n}}^{\text{drive}}=\sum_i f_i(t) \hat{n}_i\,,
\end{equation} where $f_i(t)=A_i\cos(\omega t)$, or $A_i{\rm sgn}[\sin(\omega t)]$. The co-moving Hamiltonian therefore becomes
\begin{equation}
   \calH_\text{mov}(t)=\h^0_{\widehat{n}}[\widehat{n}_i]+\h^0_{\widehat{\varphi}}[\widehat{\varphi}_i-\theta_i(t)]\,,
\end{equation}
where the $\theta_i$ is expressed as an integral over the $f_i$'s as in Eq.~\eqref{eq:thetat}, but is position-dependent now.

The first few terms in the Magnus expansion of the Floquet Hamiltonian in terms of the co-moving Hamiltonian Eq.~\eqref{eq:comov} are given by
\begin{subequations}\label{eq:magnus_all}
\begin{align}
  \calH_\text{eff}^0 &= \frac{1}{T}\int \rd t \calH_\text{mov}(t)\,, \\
  \calH_\text{eff}^1 &= \frac{1}{2iT}\int_0^T \rd t_1 \int_0^{t_1}\rd t_2 \left[\calH_\text{mov}(t_1),\calH_\text{mov}(t_2)\right]\,, \\
  \calH_\text{eff}^2 &= \frac{-1}{6T}\int_0^T \rd t_1 \int_0^{t_1}\rd t_2 \int_0^{t_2} \rd t_3\\
  &\left(\left[\calH_\text{mov}(t_1),[\calH_\text{mov}(t_2),\calH_\text{mov}(t_3)]\right]+\left(t_1\leftrightarrow t_3\right)\right) \,. \nonumber
\end{align}
\end{subequations}
The integrals can be evaluated analytically (see Supplemental Methods summarize the results below. The zeroth-order term is,
\begin{equation}
  \calH_\text{eff}^0 = \hn-\frac{E_J}{2}\sum_k\left(F_{0,k}^*e^{i\hf_k}+F_{0,k} e^{-i\hf_k}\right)\,,
\end{equation} where $F_{m,k}$ is the Fourier coefficient of $e^{i\theta_k(t)}$:
\begin{equation}\label{eq:Fm}
  e^{i\theta_k(t)}=\sum_m F_{m,k} e^{im\omega t}\,.
\end{equation} The explicit form of $F_{m,k}$ is
\begin{subequations}\label{eq:Fmk}
\begin{eqnarray}
  \text{Case I: }F_{m,k} &=& -\frac{ i A_k \omega \left(-1+(-1)^m e^{\frac{i \pi A_k}{\omega}}\right)}{\pi(A_k^2-m^2\omega^2)}\,, \\
  \text{Case II: }F_{m,k} &=& J_m\left(\frac{A_k}{\omega}\right)\,.
\end{eqnarray}
\end{subequations}
The first-order result is
\begin{equation}\label{eq:correctionfirstorder}
      \calH_\text{eff}^1=\frac{E_J}{2\omega}\sum_k\left[B_k\Delta_k \hn e^{-i\hf_k}+\tn{h.c.}\right]\,,
\end{equation}
where $\Delta_k$ is the finite difference at site $k$:
\begin{equation}
\Delta_k\hn[\hat{n}_1,\dots,\hat{n}_L]=\hn[\dots,(\hat{n}_k+1),\dots]-\hn[\dots,\hat{n}_k,\dots]\,.
\end{equation} For the $\hn$ given in Eq.\eqref{Hamiltonian1}, the explicit form is given by,
\begin{equation}\label{eq:Deltak}
    \Delta_k \hn=4 E_c (2\hat{n}_k+1)+V{\sum_{j\in \text{adj}_k}} \hat{n}_j\,,
\end{equation} where $j$ runs over the neighbors of site $k$.

The parameter $B_k$ is
\begin{equation}\label{eq:Bk}
  B_k=\sum_{m\neq 0}\frac{F_{m,k}}{m}=\left\{
                                  \begin{array}{ll}
                                    0, & \hbox{Case I;} \\
                                    \frac{\pi}{2}\mathbf{H}_0\left(\frac{A_k}{\omega}\right), & \hbox{Case II}
                                  \end{array}
                                \right.
\end{equation} where the sum ranges over all non-zero integers $m$ and $\mathbf{H}_0$ is the Struve function. Note that the first case sums to zero as $F_{m,k}/m$ is an odd function of $m$.

Finally, the second-order result is the sum of three terms $\calH_\text{eff}^2=H^2_A+H^2_B+H^2_C$, where
\begin{subequations}\label{eq:H2_main}
\beq
  H^2_A&=&\frac{E_J^2}{8\omega^2}\sum_{kl}\big[D_{kl}^*e^{i\hf_k+i\hf_l}\Delta_k \Delta_l\hn \nn\\
&&+E_{kl}e^{i\hf_k}\Delta_k \Delta_l \hn e^{-i\hf_l}+\tn{h.c.}\big]\,,\\
    H^{2}_B&=&\frac{E_J^2}{4\omega^2}\sum_{kl}\left[C_{k}^*F_{0,l}^* e^{i\hf_k+i\hf_l}\Delta_k \Delta_l\hn \right.\nn\\
&&\left.-C_{k}F_{0,l}^* e^{i\hf_k}\Delta_k \Delta_l \hn e^{-i\hf_l}+\tn{h.c.}\right]\,,\\
     H^{2}_C&=&\frac{E_J}{2\omega^2}\sum_k \left[C_k^*e^{i\hf_k}\left(\Delta_k \hn\right)^2+\tn{h.c.}\right]\,.
\eeq
\end{subequations}
The parameters $D_{kl}$, $E_{kl}$, and $C_k$ are
\begin{equation}
  D_{kl}=B_k B_l-\frac{1}{2}\sum_{m\neq 0}\frac{F_{-m,k}F_{m,l}}{m^2}\,,
\end{equation}
\begin{equation}
  E_{kl}=B_l B_k^*+\frac{1}{2}\sum_{m\neq 0}\frac{F_{m,k}^*F_{-m,l}}{m^2}\,,
\end{equation}
\begin{equation}
  C_k= \sum_{m\neq 0}\frac{F_{m,k}}{m^2}\,.
\end{equation}
Since $\Delta_k \Delta_l\hn$ is only non-zero when $k,l$ coincide, or are nearest neighbors, the second order corrections are still local Hamiltonians.

These results indicate that the Floquet-Magnus expansion is essentially a $1/\omega$ expansion, where the $A/\omega$ dependence of freezing enters via the magnitude of the coefficients of this expansion. Therefore, the expansion is expected to perform well in the high frequency limit.

Next, we consider the first-order renormalizations of the freezing point, which is especially relevant for the cosine drive in Eq.~\eqref{drive2}. We write the Floquet Hamiltonian as $\calH_\text{eff}=\hn+U$, where
\begin{equation}
\begin{split}
    U&=-\frac{E_J}{2}\sum_k \left[e^{-i\hf_k}\left(F_{0,k}-\frac{B_k}{\omega}\bar{\Delta}_k\hn\right)\right.\\
    &\left.+e^{i\hf_k}\left(F_{0,k}^*-\frac{B^*_k}{\omega}\Delta_k\hn\right)\right]\,.
\end{split}
\end{equation} Notice that due to the non-commutativity between $\hf$ and $\hat{n}$, in the first line the forward difference $\Delta_k$ has been shifted to the backward difference $\bar{\Delta}_k$ when $e^{-i\hf_k}$ is commuted to the left:
\begin{equation}
\bar{\Delta}_k\hn[\hat{n}_1,\dots,\hat{n}_L]=\hn[\dots,\hat{n}_k,\dots]-\hn[\dots,\hat{n}_k-1,\dots]\,.
\end{equation}

We consider a particular initial state $\ket{n}$ which is the eigenstate of $\hn$ and labelled by $n_i$: $\hn\ket{n}=E_n\ket{n}$, $\hat{n}_i\ket{n}=n_i\ket{n}$. To improve the freezing for state $\ket{n}$, we demand that the norm of $U\ket{n}$ should be minimized, which is achieved at
\begin{equation}\label{eq:renormalize_1}
    F_{0,k}=\frac{B_k}{2\omega}\left(\Delta_k E_n+\bar{\Delta}_k E_n\right)]\,,
\end{equation} where we have replaced $\hn$ by its eigenvalue $E_n$. Substituting the explicit expressions for the cosine drive (Eqs.\eqref{eq:Fmk}, \eqref{eq:Deltak}, \eqref{eq:Bk}), we obtain
\begin{equation}\label{eq:renormalize_1_explicit}
  J_0\left(\frac{A_k}{\omega}\right)=\frac{\pi}{2\omega}\mathbf{H}_0\left(\frac{A_k}{\omega}\right)\left(8E_c n_k+V\sum_{j\in \text{adj}_k}n_j\right)\,.
\end{equation}
Eqs.~\eqref{eq:renormalize_1} and \eqref{eq:renormalize_1_explicit} are the main result of this section, which indicates the first-order correction of the freezing point in terms of the amplitude-frequency ratio $A_k/\omega$.  Notably, the right hand side of Eq. \eqref{eq:renormalize_1_explicit} is proportional to $n_i$. Thus, the magnitude of renormalization a state experiences from the zeroth order freezing value increases with increasing $n_i$.

We notice, however, that satisfying Eq. \eqref{eq:renormalize_1} does not completely eliminate the first-order corrections. At the first-order corrected freezing point Eq. \eqref{eq:renormalize_1}, there is still a residual action of $U$ on $\ket{n}$:
\begin{equation}
    U\ket{n}=-2E_J E_c\sum_k \frac{B_k}{2\omega}\left[e^{-i\hf_k}-e^{i\hf_k}\right]\ket{n}\,.
\end{equation}  This term arises exactly due to the non-commutativity between $\Delta_k \hn=4E_c(2\hat{n}_k+1)$ and $e^{i\hf_k}$, which is the manifestation of uncertainty principle in the context of dynamical-freezing.  We also note that the first-order renormalization depends on the initial state $\ket{n}$ and can be spatially inhomogeneous. In Fig.~\ref{fig:zeroorderscaling} we test our numerical results with the theoretical prediction of the freezing under cosine drive. We compared the decay rate of $\overline{\la N(t)\ra}$ for $A/\omega$ at the zeroth-order freezing point and the first-order freezing point and performed a scaling analysis in $\omega$. In both cases the decay rate primarily scales as $1/\omega$. We observe that the decay rate of $\overline{\la N(t)\ra}$ is reduced at the first-order freezing point, in the sense that the coefficient of $1/\omega$ becomes smaller (but not vanishing). These behaviors agree with the theoretical prediction that the decay rate scales as $1/\omega$.

Next, we discuss the second-order corrections and focus on the square-wave drive Eq.~\eqref{drive1}. Since $B_k=0$ (see Eq.~\eqref{eq:Bk}), the first-order correction vanishes. Among the second-order corrections, $H_B^2\sim 1/\omega^4$ due to $F_0\sim 1/\omega^2$ and we therefore need to consider $H^2_A$ and $H^2_C$.  We note that because $H^2_C$, similar to $H^0_\text{eff}$, depends on $\hf$ only through single-site operators, it can be partially cancelled by adjusting $A_k/\omega$. The computation of the renormalized freezing point proceeds similarly to the first-order case, with the result
\begin{equation}
    F_{0,k}=\frac{C_k}{2\omega^2}\left[(\Delta_k E_n)^2+(\bar{\Delta}_k E_n)^2\right]\,.
\end{equation} Subsituting the parameters for the square-wave drive, the explicit equation for $\alpha_k=A_k/\omega$ for the renormalized freezing point is
\begin{equation}
\begin{split}
    &\alpha_k^2(e^{i\pi\alpha_k}-1)=-\frac{(\Delta_k E_n)^2+(\bar{\Delta}_k E_n)^2}{12\omega^2}\\
    &\times\left[-6-6i\pi\alpha_k+2\pi^2\alpha_k^2+e^{i\pi\alpha_k}(6+\pi^2\alpha_k^2)\right]\,,
\end{split}
\end{equation} where
\begin{equation}
    \Delta_k E_n=4E_c(2n_k+1)+V\sum_{j\in \text{adj}_k} n_j\,,
\end{equation}
\begin{equation}
    \bar{\Delta}_k E_n=4E_c(2n_k-1)+V\sum_{j\in \text{adj}_k} n_j\,.
\end{equation}
Again, this does not completely cancel the second order correction due to non-commutativity.

Furthermore, at the freezing point $H^2_A\neq 0$, and it contains terms of the form $e^{i\hat{\varphi}_k\pm i\hat{\varphi}_l}$, which cannot be canceled by adjusting $A_k/\omega$. Therefore, both $H^2_A$ and $H^2_C$ will contribute to non-conservation at the freezing point, and we  expect the residual non-conservation rate to scale as $1/\omega^2$ for the system under square-wave drive.

\section*{Discussion}
\label{outlook}

In this work, we have analyzed the phenomenon of dynamical freezing using a variety of complementary perspectives for a paradigmatic many-body system of coupled Josephson junctions. Freezing is demonstrated in an interacting translationally invariant system, and can be understood within an effective Floquet-Magnus high-frequency expansion. One of our most interesting observations is the semiclassical limit of dynamical freezing, where the solutions for the coupled Hamilton's equations of motion show a dramatic reduction of their chaotic properties at the frozen points. However, it is also clear that freezing is not exact, and we have been able to address both the renormalization of the freezing point and the residual slow dynamics due to the higher-order corrections within the same Floquet-Magnus expansion. Importantly, the structure of this expansion is dependent on the drive, making some drive protocols more resilient to the emergent conservation laws compared to others (e.g. square-wave vs. cosine-wave drive). Given our refined understanding of the structure of these corrections, it will be an interesting future exercise to consider more complex Floquet drive protocols that can diminish their effects.

Clearly, one practical application of this protocol will be to adapt it in a fashion where one is able to protect encoded information up to late times. In the current transmon-based setting, the key impediment towards this is due to the noisy offset charge \cite{PhysRevLett.108.240502,PhysRevLett.117.190503}, which has been ignored for simplicity in our analysis.
At the freezing point, where the Hamiltonian effectively becomes $\h^0_{\widehat{n}}$, the spatio-temporal dependence of the offset charge can play a significant role in the decoherence. Finding alternative routes to mitigate this effect in a realistic superconducting device based architecture remains an exciting open direction.

\section*{Methods}
\label{prelim}
Here we introduce the basics of Floquet theory that underpins the theory throughout this work and introduce the correspondence between the quantum and semi-classical limits of the coupled transmon system.

\subsection*{Floquet Theory}
\customlabel{sec:FM}{(\textbf{Floquet Theory})}
Our time-dependent Hamiltonian $H(t) = H_0 + H_{\rm{D}}(t)$ consists of two parts --- the static piece, $H_{0}\equiv{\cal{H}}_{\widehat{n}}^0 + {\cal{H}}_{\widehat{\varphi}}^0$, and the driven piece, $H_{\rm{D}}(t)\equiv f(t)\h^{\rm{drive}}_{\widehat{n}}$.
Due to the discrete time translational symmetry associated with the period $T\equiv 2\pi/ \omega$, the solutions of the Schr\"odinger equation can be written as $\ket{\psi_{\alpha}(t)}=e^{-i\epsilon_{\alpha}t}\ket{u_{\alpha}(t)}$, with time-periodic Floquet wavefunction, $\ket{u_{\alpha}(t+T)}=\ket{u_{\alpha}(t)}$, and Floquet quasi-energy, $\epsilon_{\alpha} \in (-\omega/2,\omega/2]$, respectively. The Floquet-Schr\"odinger equation is written as~\cite{eckardt2015high,bukov2015universal},
\begin{equation}
    [H(t)-i\partial_{t}]\ket{u_{\alpha}(t)}=\epsilon_{\alpha}\ket{u_{\alpha}(t)}.
\end{equation}
The Floquet eigenspectrum can be obtained by diagonalizing the effective Floquet Hamiltonian, $H_{\rm{eff}}$, such that
\begin{equation}
 \begin{split}
    U(T,0)\ket{u_{\alpha}(0)}&=\exp{[-iH_{\rm{eff}}]}\ket{u_{\alpha}(0)}\\
    &=\exp{(-i\epsilon_{\alpha}T)}\ket{u_{\alpha}(0)},
 \end{split}
\end{equation}
where $U(t,t')$ is the usual time evolution operator $U(t,t')\ket{\psi (t')}=\ket{\psi (t)}$,
\begin{equation}
    U(t,t')=\mathcal{T} \exp{\Big(-i \int_{t'}^{t} dt'' H(t'')    \Big)},
\end{equation}
and $\mathcal{T}$ takes time-ordering into account. We are interested in the late time behavior of the system, and not on the detailed time dependence within individual periods. Hence, we will be primarily interested in the stroboscopic evolution of various observables which can be described by the effective Floquet Hamiltonian $H_{\rm eff}$.

In the limit of both large driving amplitude and frequency {  (relative to the transmon frequency given in Eq. \eqref{eq:transmon_frequency})}, $H_{\rm eff}$ can be obtained from the modified magnus expansion \cite{freezing4}, which we review now. We consider moving into the co-moving frame of the driving term, via the unitary transform
\begin{equation}\label{eq:thetat}
    W(t)=\exp(-i\theta(t)\h_{\widehat{n}}^\text{drive})\,,
\end{equation} where $\theta(t)=\int_0^t \rd t' f(t')$ with the explicit form
\begin{subequations}
\begin{eqnarray}
    \text{Case I: } \theta(t)&=& \frac{2\pi A}{\omega}\left[1/2-\left|\{\omega t/(2\pi)\}-1/2\right|\right]\,,\\
    \text{Case II: } \theta(t)&=&\frac{A}{\omega}\sin\omega t\,.
\end{eqnarray}
\end{subequations} Here $\{x\}$ means taking the decimal parts of $x$.
 The co-moving Hamiltonian is given by
\beq\label{eq:comov}
  \calH_\text{mov}(t)&=&W(t)^\dagger \left[H(t)-i\partial_t\right] W(t)\nn\\
  &=&W(t)^\dagger (\h^0_{\widehat{n}}+\h^0_{\widehat{\varphi}}) W(t)\,,
\eeq which is nominally independent of the large driving term $\h_{\widehat{n}}^\text{drive}$. The co-moving Hamiltonian $\h_\text{mov}$ can be evaluated using the Baker-Campell-Hausdorff formula, and the result is:
\begin{equation}
    \calH_\text{mov}(t)=\h^0_{\widehat{n}}[\widehat{n}_i]+\h^0_{\widehat{\varphi}}[\widehat{\varphi}_i-\theta(t)]\,.
\end{equation}
{  Here we use the notation $\h^0_{\widehat{n}}[\widehat{n}_i]$ to indicate that $\h^0_{\widehat{n}}$ is being written as a function the operator $\widehat{n}_i$ and similarly for other Hamiltonians throughout the paper.} The Magnus expansion \cite{freezing4} states that the leading order Floquet effective Hamiltonian is given by the one-period average of $\h_\text{mov}(t)$:
\begin{equation}
    \h_\text{eff}^0=\frac{1}{T}\int_0^T \rd t~ \h_\text{mov}(t)\,.
\end{equation} For the model under consideration, we find
\begin{subequations}
\begin{eqnarray}
        \h_\text{eff}^0 &=& \h^0_{\widehat{n}}[\widehat{n}_i]-\sum_i\frac{\omega E_J}{ \pi A}\left[\sin(\widehat{\varphi}_i)+\sin(A\pi/\omega-\widehat{\varphi}_i)\right] + ... \,,\nonumber \label{nonintcond1} \\
        \\
        \h_\text{eff}^0 &=& \h^0_{\widehat{n}}[\widehat{n}_i]- E_J J_0\left(\frac{A}{\omega}\right) \sum_i\cos(\widehat{\varphi}_i) + ...\, \label{nonintcond2}.
\end{eqnarray}
\end{subequations} Here $\dots$ denote higher order corrections in $1/\omega$ which will be studied in further detail in Sec.~\ref{results:FM}; see also Supplemental Methods \ref{app:FM}.
Therefore, we find at a set of isolated points, the non-commuting $\varphi$-dependent terms vanish, and the effective Hamiltonian becomes seemingly integrable (at leading order):
\begin{subequations}\label{eq:freezingpoint0}
\beq
{\rm{Case~ I:}}&&~~\frac{A}{\omega}\approx 2n, \ {n\in \mathbb{Z}}\label{condn1},\\
{\rm{Case~ II:}}&&~~\frac{A}{\omega} \approx \text{zeros of Bessel function} \ J_{0}.\label{condn2}
\eeq
\end{subequations}
Given the structure of freezing and the form of the underlying drive Hamiltonian, we expect that the quantity $\widehat{N}= (1/L) \sum_{i}\widehat{n}_{i}$, or any operator that commutes with $\widehat{N}$ develops an emergent conservation law. The extent of thermalization can become extremely slow at the freezing points, in marked contrast to the rapid relaxation away from freezing.

\subsection*{Semi-Classical Limit}
\customlabel{sec:SClimit}{\textbf{(Semi-Classical Limit)}}
The dynamics of the coupled transmons is amenable to a semi-classical analysis, where we replace the non-commuting operators $\widehat\varphi$ and $\widehat{n}$ by classical real variables, and treating the commutator bracket as a Poisson bracket, $\{\varphi_{i},n_{j}\}=\delta_{ij}$, respectively. Recall that in the absence of a Floquet drive, the semi-classical dynamics of transmon arrays has been analyzed, and the behavior is generically expected to be chaotic~\cite{PRXQuantum.4.020312,borner2023classical} without substantial frequency detuning or tunable coupling~\cite{berke2022transmon}. Our goal will be to study how the Floquet drive in the absence of any other strategy helps arrest the growth of chaos, and protect the memory associated with generic initial states.

The Hamilton equations of motion for the coupled set of phase and number variables in Eq.~\eqref{Eq:totaHam} are given by,
\begin{subequations}
\beq
    \dot{\varphi}_{i} &=& \{\varphi_{i},\h\}=\dfrac{\partial \h}{\partial n_{i}}=8E_{c}n_{i}+V\sum_{j\in \text{adj}_i}n_{j}+f(t), \label{SC1}\\
    \dot{n}_{i} &=& \{n_{i},\h\}=-\dfrac{\partial \h}{\partial \varphi_{i}}=-E_{J_{i}}\sin{\varphi_{i}} \label{SC2},
\eeq
\end{subequations}
where $\sum_{j\in \text{adj}_i}$ extends over the nearest neighbors of $i$. In the language of pendula the interaction between the transmons is a momentum-momentum interaction that is due to the capacitive coupling. In addition to obtaining the full dynamics associated with $\vp_i(t),~n_i(t)$, we will also  diagnose chaos by computing the Lyapunov exponent~\cite{strogatz2018nonlinear}. Specifically, we focus on the maximal Lyapunov exponent (MLE), $\lambda_{L}$, which determines the rate at which trajectories that are initially ``close" diverge with time, $\delta \pi(t) \approx \delta \pi \exp{(\lambda_L t)}$.{   We define $\lambda_L$ as follows~\cite{vulpiani2009chaos}. Consider two trajectories in phase space with an infinitesimal initial separation $\delta\pi_0$ at time $t=0$, and let $\delta\pi(t)$ be the resulting separation at a later time $t$. Then
\begin{align}
    \lambda_L = \lim_{t\rightarrow \infty}\lim_{|\delta\pi_0|\rightarrow 0}\frac{1}{t}\log\left(\frac{|\delta\pi(t)|}{|\delta\pi_0|}\right)\label{eq:lyapunov_def}.
\end{align}}

By contrasting the Lyapunov exponent for the static vs. driven transmon arrays, we highlight in Results~\ref{results:SC} how dynamical freezing effectively arrests the growth of chaos at special values of $A/\omega$.

\section*{Resource Availability}

\subsection*{Lead Contact}
Further information and requests should be directed to and will be fulfilled by Debanjan Chowdhury (debanjanchowdhury@cornell.edu)

\subsection*{Materials availability}
This study did not generate new materials.

\subsection*{Data and code availability}
\begin{itemize}
    \item All the theoretical data generated in this study has been deposited at \href{https://doi.org/10.5281/zenodo.14594541}{Zenodo} and is publicly available as of the date of publication.
    \item The custom computer codes used to generate the results reported in this paper are available from the lead contact upon request.
\end{itemize}

\section*{Acknowledgments}

We thank S.D. Börner, A. Das, V. Fatemi, B. Kobrin, Y. Lensky, R. Moessner, E. Rosenberg, P. Roushan, S. Roy, V. Smelyanskiy, C. Tahan, G. Vidal and B. Ware for a number of useful discussions leading up to this work. RM thanks M. Bukov and S. Bandyopadhyay for many useful suggestions regarding QuSpin PYTHON package~\cite{weinberg2017quspin,weinberg2019quspin}, and thanks S.D. Börner for providing the Julia code dynamical.jl~\cite{datseris2018dynamicalsystems} for solving the semiclassical equations for the transmon arrays.  RM is supported by a Fulbright-Nehru Grant No. 2877/FNDR/2023-2024 sponsored by the Bureau of Educational and Cultural Affairs of the United States Department of State. HG is supported by a Wilkins postdoctoral fellowship at Cornell University. KL acknowledges that this material is based upon work supported by the National Science Foundation Graduate Research Fellowship under Grant No. DGE – 2139899. DC is supported in part by a New Frontier Grant awarded by the College of Arts and Sciences at Cornell University and by a Sloan research fellowship from the Alfred P. Sloan foundation.

\section*{Author Contributions}
This research was conceived and supervised by DC. RM performed all of the exact diagonalization computations. HG performed the Floquet-Magnus computations. KL performed the semiclassical computations. All the authors analyzed the results and wrote the paper.

\section*{Declaration of Interests}
The authors declare no competing interests.

\section*{MAIN FIGURE TITLES AND LEGENDS}

\noindent\includegraphics[width=0.85\linewidth]{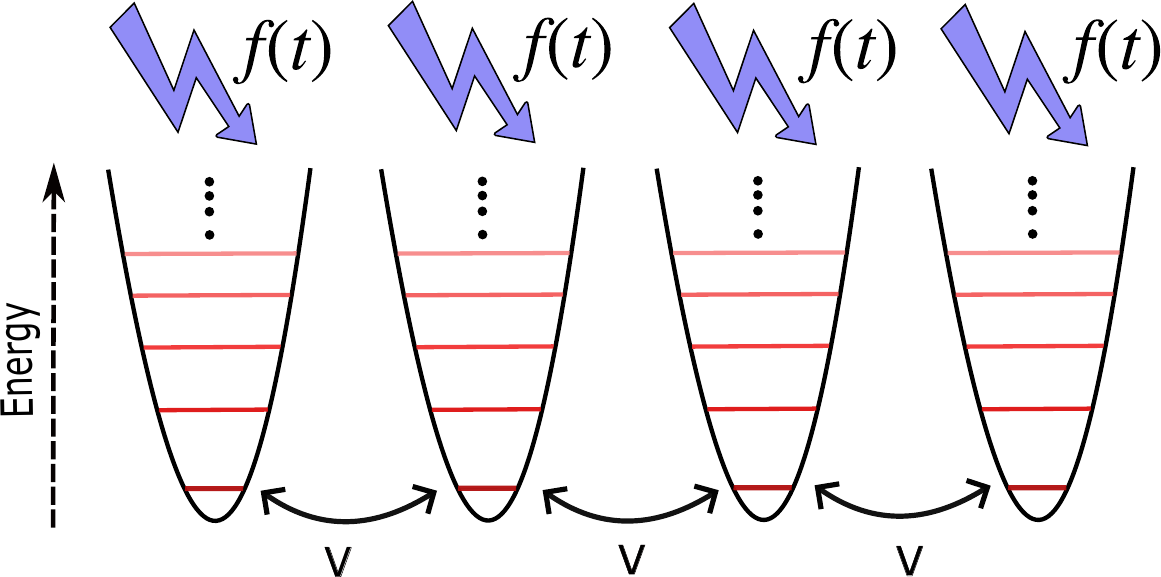}

\subsection*{Figure 1. Schematic representation of a transmon array}
\customlabel{fig:Schemetic}{1}

The energy levels (red lines) for non-linear transmons with an unbounded spectrum coupled via capacitive interactions. The external Floquet drive, $f(t)$, couples capacitively to each transmon. $V$ is the capacitive coupling between the transmons.

\noindent\includegraphics[width=0.85\linewidth]{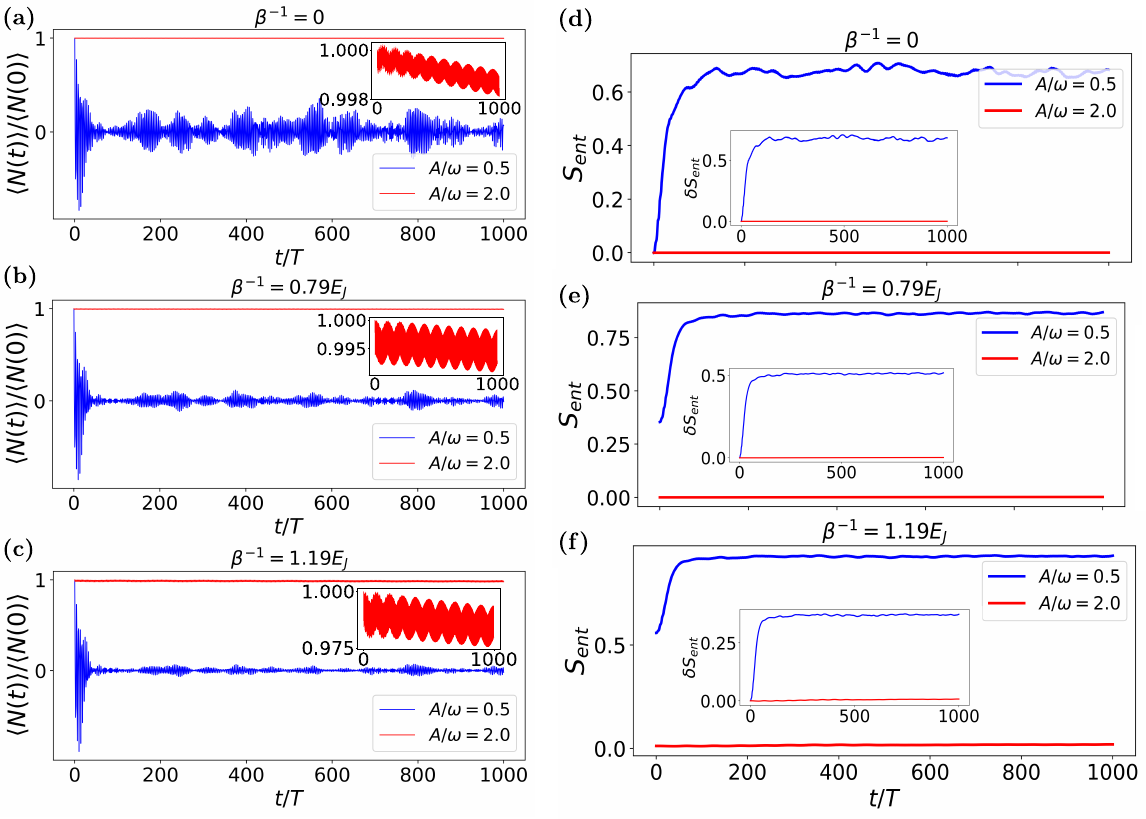}

\subsection*{Figure 2. Emergent conservation and entanglement entropy for the square-wave drive}
\customlabel{fig:ndtstrobo}{2}
In (a), (b), and (c) we show the stroboscopic time evolution of normalized $\widehat{N}(t)$ for different driving strengths for the square drive in Eq.~\eqref{drive1}. The{  one-dimensional} system size is $L=8$, with an on-site Hilbert space dimension $\mathcal{D}=3$. The drive frequency {  $\omega/(2\pi) =9.32$ GHz}, and  other parameters are as introduced in the main text. The initial state corresponds to the: (a) ground state, or thermal pure state with (b) $\beta^{-1}=0.79E_{J}$, (c) $\beta^{-1}=1.19E_{J}$, of $H(0)$. At freezing, there is a very slow decay of the normalized $\la\widehat{N}(t)\ra$ over time (inset). In (d), (e), and (f) we show the stroboscopic evolution of half-chain entanglement entropy at (red) and away from (blue) freezing for a square-wave drive for the same initial thermal states and same parameters as above. In the inset, we plot the relative change in entropy. At the freezing point, the entanglement growth is sufficiently suppressed, whereas, away from freezing, it increases rapidly before eventually saturating.


\noindent\includegraphics[width=0.85\linewidth]{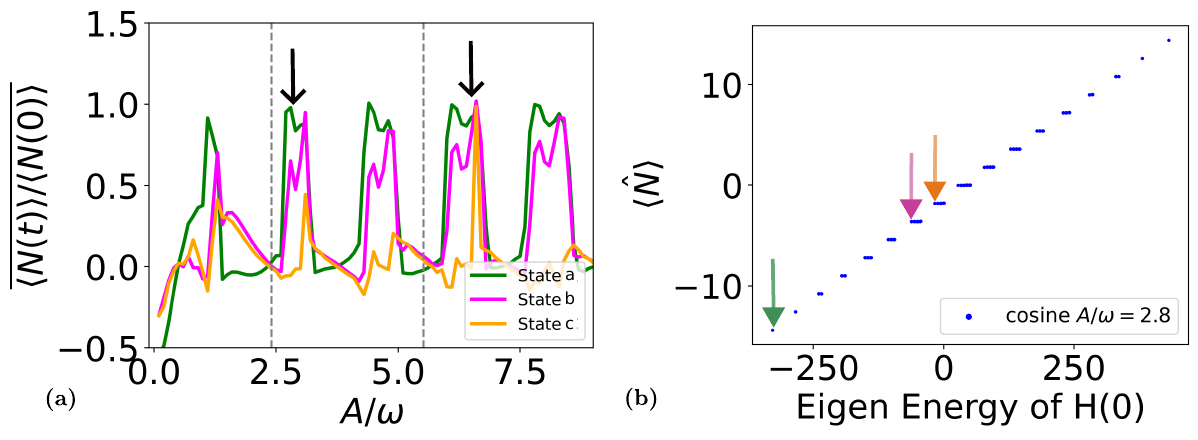}

\subsection*{Figure 3. Renormalization of freezing points}
\customlabel{fig:cosonefidelity}{3}
(a) Time-averaged ${\cal{F}}$ (averaged over 100 Floquet cycles) as a function of $A/ \omega$ for the cosine drive. We choose $\ket{\psi(0)}$ from the eigenspectrum of $H(0)$. Primary freezing points,  indicated by black arrows, shift from the zeroth-order prediction of freezing point in Eq.~\eqref{condn2} {  (dashed lines)}. The shift is due to the higher order corrections in the Magnus expansion, discussed in Sec.~\ref{results:FM}.{  (b) The expectation value of the operator $\hat{N}$ is calculated with respect to the specific eigenstates of $H(0)$ and plotted against the corresponding eigen-energy. The states chosen to compute $\la\widehat{N}(t)\ra$ are indicated with the colored arrows and labeled $a,~b,~c$, respectively.

\noindent\includegraphics[width=0.85\linewidth]{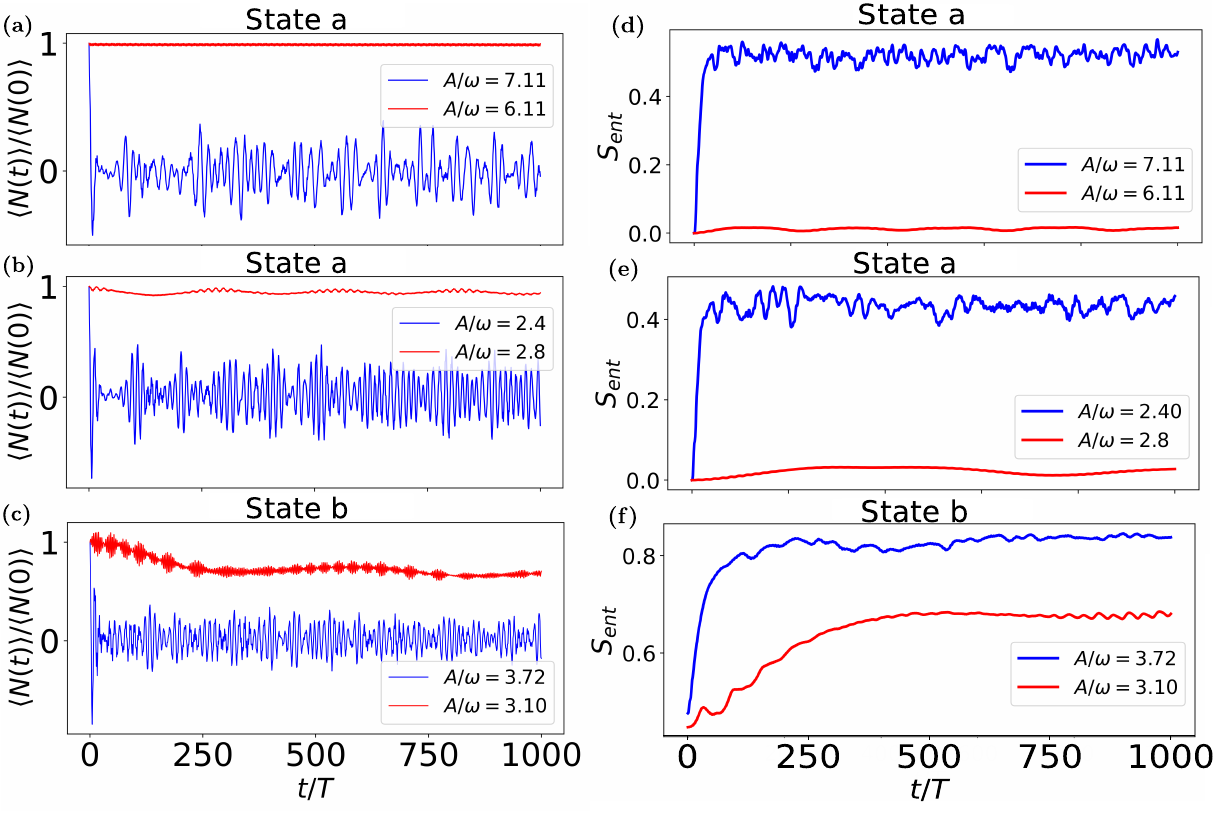}

\subsection*{Figure 4. Emergent conservation and entanglement entropy for the cosine drive}
\customlabel{fig:cosinendt}{4}
In (a), (b), and (c) we show the stroboscopic time evolution of normalized $\widehat{N}(t)$ for the cosine drive given by Eq.\eqref{drive2}. The one-dimensional system size is $L=8$, with an on-site Hilbert space dimension $\mathcal{D}=3$. The drive frequency $\omega/(2\pi) =9.32$ GHz, and  other parameters are as introduced in the main text. The initial state are chosen from the eigenstates of $H(0)$. The renormalized freezing points are obtained from Fig.~\ref{fig:cosonefidelity}.{  At the \textit{renormalized freezing point}, $\widehat{N}(t)$ becomes approximately conserved. Interestingly, in (b), the system rapidly loses its memory at $A/ \omega$ = 2.4, the freezing point predicted by the zeroth-order Magnus expansion, that suggest significant contributions from the higher order corrections. In (d), (e), and (f) we show the stroboscopic evolution of half-chain entanglement entropy at (red), and away from (blue) freezing for a cosine wave drive for the same parameters and initial states, $\ket{\psi(0)}$, as above. Even at the renormalized freezing point, we observe a significant growth in EE compared to the square wave drive, which can be attributed to higher-order corrections, even in the first-order Magnus expansion (see Sec.~\ref{results:FM}).

\noindent\includegraphics[width=0.85\linewidth]{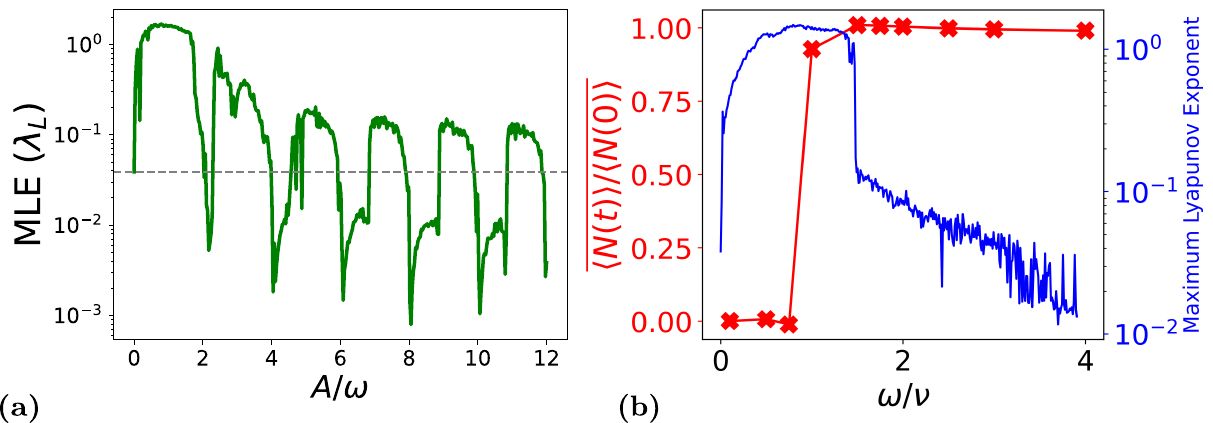}

\subsection*{Figure 5. Phenomenology of semi-classical freezing}
\customlabel{fig:classicalLE}{5}
The maximum Lyapunov exponent (MLE), $\lambda_L$, as a function of $A/\omega$ for a fixed drive frequency ({  $\omega/(2\pi) = 15$ GHz}) with the square-wave drive,{  for a length $8$ one-dimensional chain} initialized in the $1010...$ state. The strong suppression of $\lambda_L$ coincides with the location of freezing points in Eq.~\ref{condn1}. Dashed horizontal line shows $\lambda_L$ for the static hamiltonian. (b) Correspondence between quantum and classical dynamics as a function of the ratio of drive frequency to transmon frequency{  (Defined in Eq.~\eqref{eq:transmon_frequency}; see also Ref.~\cite{berke2022transmon} for a mapping between the transmon and Bose-Hubbard hamiltonian in the context of defining these frequencies}) at freezing with $A/\omega=2$ kept fixed. The quantum expectation $\langle \widehat{N}(t)\rangle$ is computed by time-averaging over 100 Floquet cycles in the ground state of $H(0)$ which is plotted as the red curve. The blue curve is the corresponding maximum Lypunov exponent. {For the quantum and semi-classical systems there exist similar thresholds for $\omega$ below which freezing no longer occurs.} For both plots $L=8$, $E_c=330$ MHz,  $E_J=12.58$ GHz , and $V=50$ MHz.


\noindent\includegraphics[width=1.0\linewidth]{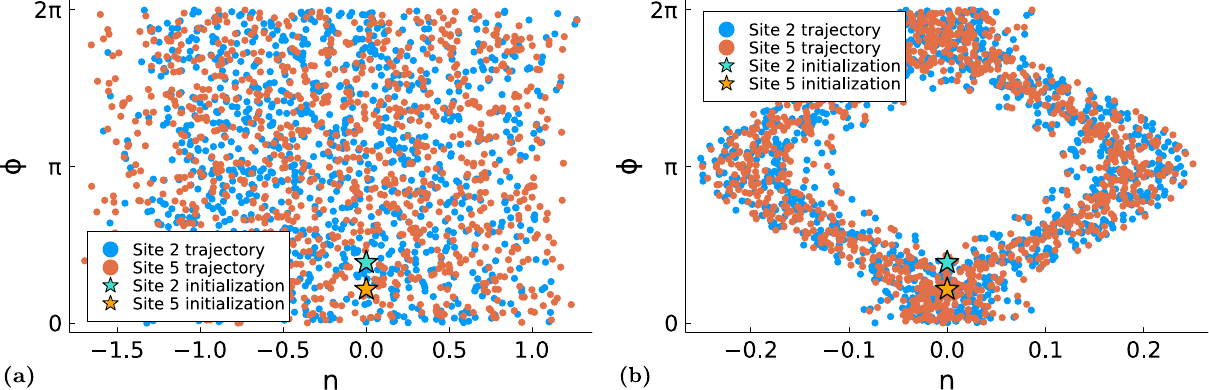}

\subsection*{Figure 6. Phase space behavior of semi-classical freezing}
\customlabel{fig:phasespace8}{6}

Phase-space trajectories for two transmons (site $2$ and $5$, respectively) for a one-dimensional array with $L=8$ transmons, for initial state $1010...$ and a square-wave drive with {  $\omega/(2\pi)=15$ GHz} over $250$ cycles.{  The ratios of drive amplitude to frequency are (a) $A/\omega = 7.3$, and (b) $A/\omega = 8$ respectively. Note that away from freezing, the values of $n$ vary ergodically, while at freezing it remains tightly constrained around its initial value of $0$.

\noindent\includegraphics[width=0.75\linewidth]{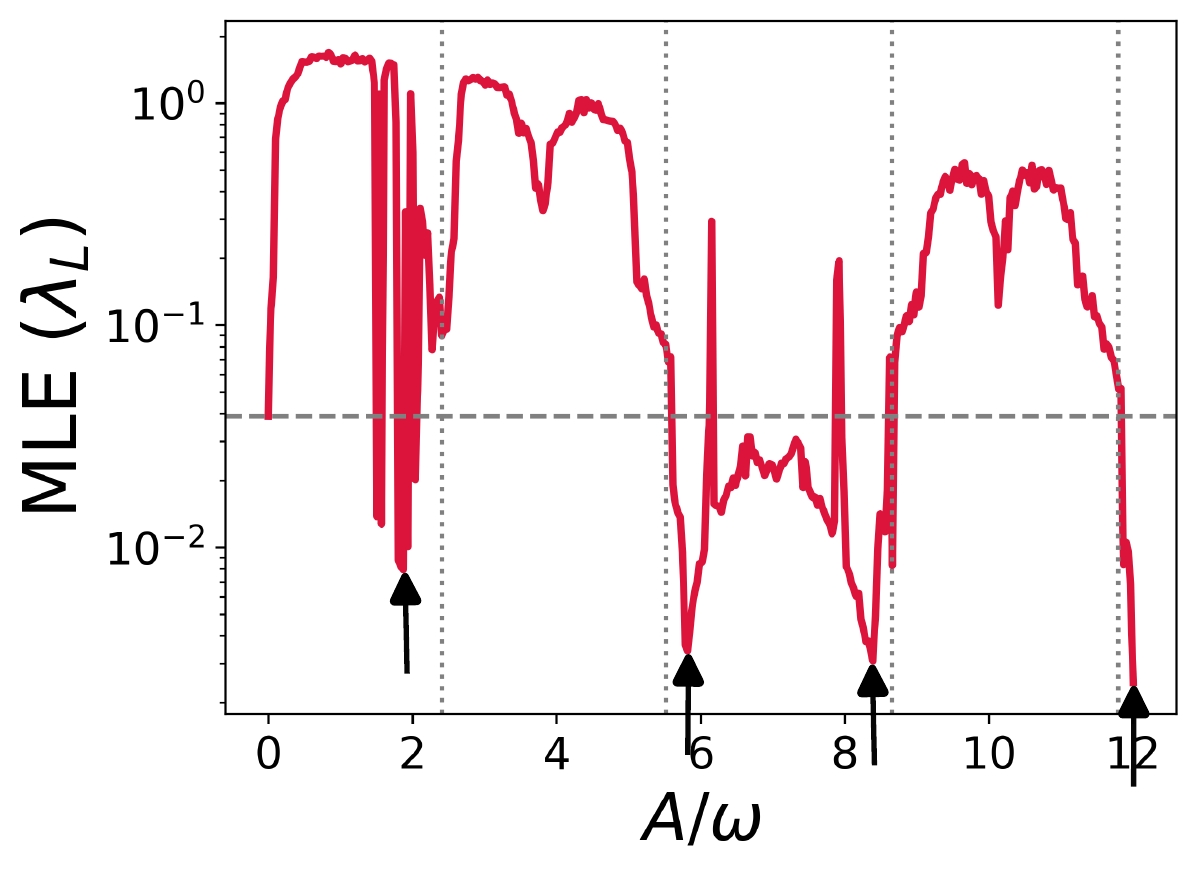}

\subsection*{Figure 7. Semi-classical freezing for the cosine drive}
\customlabel{Lyapnuov_cosinedrive}{7}

The maximum Lyapunov exponent as a function of $A/\omega$ for the cosine drive. The system with $L=8$ transmons is initialized in an $1010...$ state with a drive frequency {  $\omega/(2\pi)=15$ GHz}. The freezing points are indicated by arrows. The horizontal dashed line indicates $\lambda_L$ for the static Hamiltonian. The vertical dotted lines indicate the freezing points associated with the zeroth-order term in Eq.~\ref{condn2}.


\noindent\includegraphics[width=0.85\linewidth]{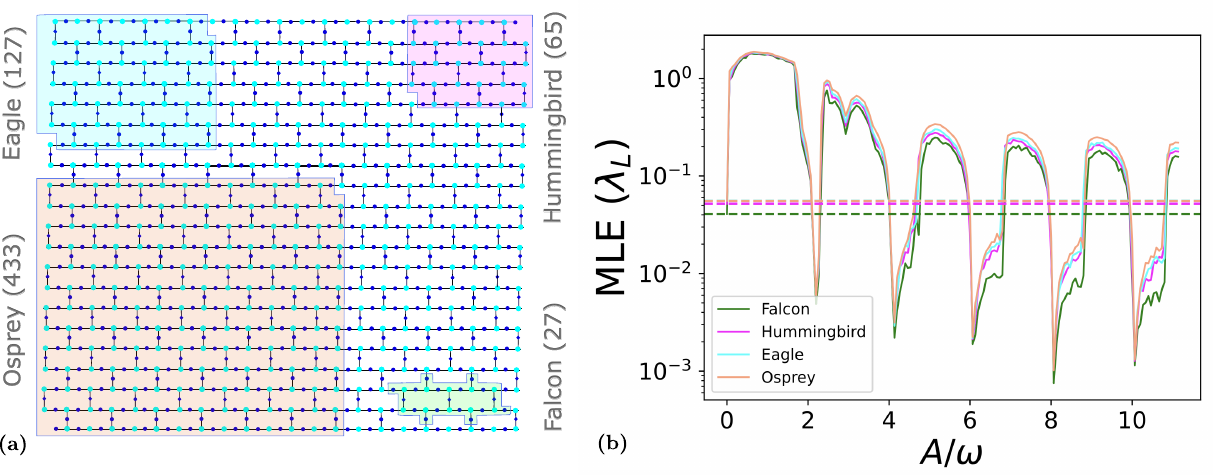}

\subsection*{Figure 8. Semi-classical freezing for IBM's lattice geometries}
\customlabel{fig:IBM_simulations}{8}

(a) Heavy hexagon geometries used in IBM's quantum processors. The light and dark blue qubits arranged in a checkerboard pattern are initialized as $1$ and $0$, respectively. The colored regions show the geometries of IBM's Falcon (green, $26$ qubits), Hummingbird (pink, $65$ qubits), Eagle (teal, $127$ qubits), and Osprey (orange, $433$ qubits) processors. (b) MLE as a function of $A/\omega$ for the different geometries with up to $433$ semi-classical qubits. All systems are driven at a fixed frequency ({  $\omega/(2\pi)=15$ GHz}) using the square-wave drive.


\noindent\includegraphics[width=0.75\linewidth]{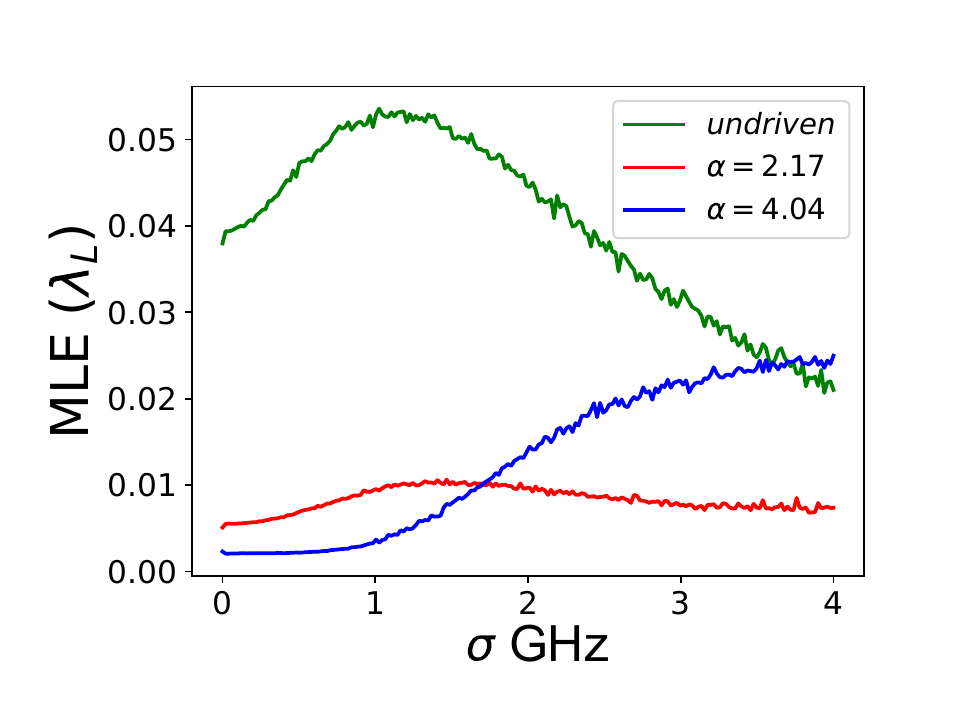}

\subsection*{Figure 9. Dependency of semi-classical freezing on disorder}
\customlabel{fig:Disorder_averaging}{9}

The maximum Lyapunov exponent as a function of disorder strength at the first and second renormalized freezing point, compared to the undriven case. Each curve is obtained for a one-dimensional array of length $8$ initialized in the $0101\ldots$ state and driven with a $\omega/(2\pi)=15$ GHz square-wave drive. Disorder is implemented by choosing $E_{J_i}\sim\mathcal{N}(12.58, \sigma^2)$ GHz (normal distribution) independently for each site in the chain. Results are averaged over $1000$ disorder realizations. Typical disorder in IBM devices is on the order of $\sigma\approx 0.7$ GHz \cite{IBM}.


\noindent\includegraphics[width=0.85\linewidth]{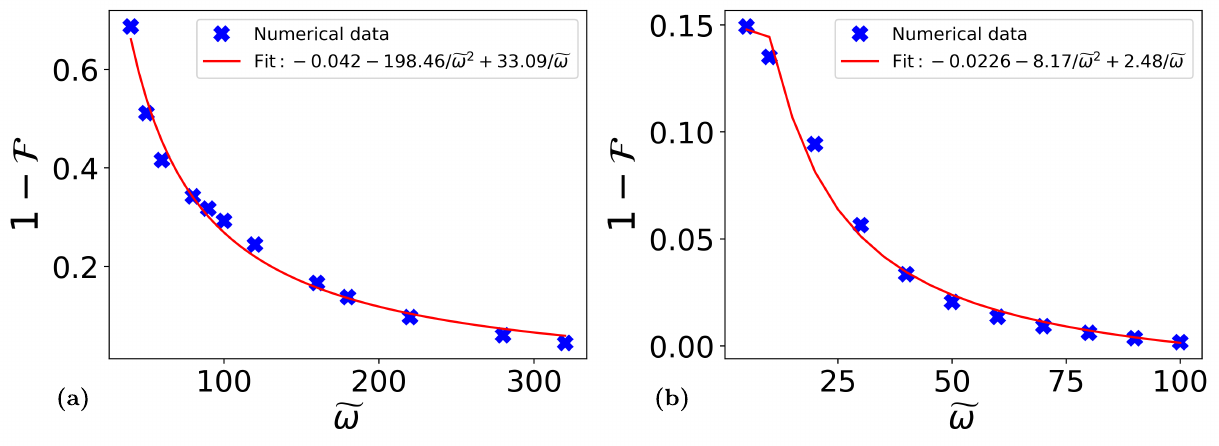}

\subsection*{Figure 10. Dependency of freezing strength on drive frequency}
\customlabel{fig:zeroorderscaling}{10}

(a) The time-averaged observable, ${\cal{F}}$ (Eq.~\eqref{Eq.Fidelity}),   as a function of $\widetilde{\omega} = \omega/\nu$ for $\nu$ given in Eq. \eqref{eq:transmon_frequency}. We choose $\ket{\psi(0)}$ as the ground state of $H(0)$ and fix $A/\omega=2.4048$ at the first freezing point within the zeroth-order Magnus expansion. The scaling is predominantly  proportional to $1/ \omega$ (Eq.~\eqref{eq:correctionfirstorder}). (b) ${\cal{F}}$ at the first renormalized freezing value of $A/\omega$ (taking into account the zeroth and first-order corrections). The other parameters are the same as in (a).


\newpage

\appendix

\title{Supplementary information for ``Arresting Quantum Chaos Dynamically in Transmon Arrays"}

\maketitle

\renewcommand{\theequation}{S.\arabic{equation}} 
\setcounter{equation}{0} 
\renewcommand{\thefigure}{S.\arabic{figure}}

\section{{  Freezing for Large Hilbert-Space Dimension}}
\label{App:hilbert_dim}

{ Here we present additional results for larger local Hilbert space dimensions, in order to fully capture the relevant freezing phenomenology. To account for the chaotic effects of the driven transmon, we must truncate the local Hilbert space such that all of the energy levels extending to the top of the cosine potential are included. The approximate eigen-energy of the single transmon is  given by
\begin{equation}
    \epsilon_{n}=\sqrt{8E_{J}E_{c}}(n+\frac{1}{2})-\frac{E_{c}}{12}(6n^2+6n+3).
\end{equation}
For our choice of parameters, $E_{J}=12.58$ GHz and $E_{c}=330$ MHz, the number of bound states corresponding to the maximum number $n_{\text{max}}$ such that the $\epsilon_{n_{\text{max}}}$ remains below the barrier height, is approximately given by $E_{J}$. The number of bound state in our case is thus $N_{\text{bound}} \approx 6$. Results for local Hilbert space dimension ${\cal{H}}_D=8$ for a one-dimensional chain of $4$ transmons are shown in Fig.~\ref{fig:firstorderscaling}. To correctly approximate the cosine potential, we expand the potential to $8^{\text{th}}$ order in $\phi$ as
\begin{equation}
    \cos{(\hat{\phi})}=I-\frac{\hat{\phi}^2}{2!}+\frac{\hat{\phi}^4}{4!}-\frac{\hat{\phi}^6}{6!}+\frac{\hat{\phi}^8}{8!}.
\end{equation} We find that freezing still occurs near the $A/\omega=2$ zeroth-order freezing point.}

{ In Fig.\ref{fig:HilbertspaceSystemsize}, we systematically study the freezing phenomenology as a function of both system size and onsite Hilbert space dimension for the square wave drive. We find that the freezing point is robust. In (a), we observe that the expectation value of $N(t)/N(0)$ converges as a function of system size when the onsite Hilbert space dimension is kept fixed. On the other hand, when the Hilbert space dimension is increased while keeping the system size fixed (as shown in Fig.~\ref{fig:HilbertspaceSystemsize} (b)), the freezing point is slightly shifted away from the predictions of the leading-order Magnus expansion.

Away from the freezing point, there exist variations of $N(t)/N(0)$ concerning the local Hilbert space dimension $\mathcal{D}$. This phenomenon is anticipated since, away from the freezing point, the system rapidly thermalizes and enters a chaotic regime, as evidenced by the entanglement entropy depicted in Fig.~\ref{fig:ndtstrobo}(d),(e),(f) in the main text. The saturation value of the entanglement entropy is expected to sensitively depend on $\mathcal{D}$, and similar occurrences are expected in $N(t)/N(0)$. However, at the freezing point, due to the emergent conservation and integrability, the state remains localized in the Fock basis, and is therefore independent of $\mathcal{D}$.

\section{Analytical details of the Floquet-Magnus Expansion}
\label{app:FM}
Here we provide additional details regarding the Floquet-Magnus expansion.

\subsection{General Expression for the Floquet Hamiltonian}

Continuing the discussion in the main text, we work in the co-moving frame of the drive Hamiltonian, where the time-evolution unitary is given by
\begin{equation}
    U_I(t)={\mathcal T} \exp\left(-i\int_0^t\rd t' \calH_\text{mov}(t')\right)\,,
\end{equation} where $\mathcal{T}$ denotes time ordering. $U_I(t)$ satisfies the Schr\"{o}dinger equation in the interaction picture
\begin{equation}\label{eq:SchUI}
    i\partial_t U_I(t)=\calH_\text{mov}(t) U_I(t)\,.
\end{equation}
We can write $U_I(t)=\exp[\Omega(t)]$, where the Floquet Hamiltonian will be identified with $\calH_\text{eff}=i\Omega(T)/T$. We can rewrite the derivative of $\partial_t U_I$ using the identity
\begin{equation}
\begin{split}
    \partial_t e^{\Omega(t)}&=\int_0^1 \rd \lambda e^{\lambda\Omega(t)} \partial_t \Omega(t) e^{(1-\lambda)\Omega(t)}\\
    &=h({\rm ad}_{\Omega(t)}) \partial_t \Omega e^{\Omega(t)}\,,
\end{split}
\end{equation} where $h(x)=(e^x-1)/x$ and ${\rm ad}_X Y=[X,Y]$. Combining this with Eq.~\eqref{eq:SchUI}, we obtain,
\begin{equation}
    \partial_t \Omega(t)=-i h\left({\rm ad}_{\Omega(t)}\right) \calH_\text{mov}(t)\,.
\end{equation} This equation can be integrated iteratively in powers of $\calH_\text{mov}$ with boundary condition $\Omega(0)=0$, and we obtain the first few terms in the Magnus-Floquet expansion for $\calH_\text{eff}$:
\begin{align}
  \calH_\text{eff}^0 &= \frac{1}{T}\int_0^T \rd t \calH_\text{mov}(t)\,, \label{eq:H0_app} \\
  \calH_\text{eff}^1 &= \frac{1}{2iT}\int_0^T \rd t_1 \int_0^{t_1}\rd t_2 \left[\calH_\text{mov}(t_1),\calH_\text{mov}(t_2)\right]\,,\label{eq:H1_app} \\
  \calH_\text{eff}^2 &= \frac{-1}{6T}\int_0^T \rd t_1 \int_0^{t_1}\rd t_2 \int_0^{t_2} \rd t_3 \label{eq:H2_app}\\
  &\left(\left[\calH_\text{mov}(t_1),[\calH_\text{mov}(t_2),\calH_\text{mov}(t_3)]\right]+\left(t_1\leftrightarrow t_3\right)\right) \,. \nonumber
\end{align}

\subsection{Magnus Expansion of the Transmon Hamiltonian}

In this section we explicitly evaluate Eqs.~\eqref{eq:H0_app}-\eqref{eq:H2_app} for the transmon Hamiltonian. The co-moving Hamiltonian is given by
\begin{equation}
    \calH_\text{mov}(t)=\hn-E_J \sum_k \cos(\hf_k-\theta_k(t))\,,
\end{equation} where
\begin{subequations}
\begin{eqnarray}
    \text{Case I: } \theta(t)&=& \frac{2\pi A_k}{\omega}\left[1/2-\left|\{\omega t/(2\pi)\}-1/2\right|\right]\,,\\
    \text{Case II: } \theta(t)&=&\frac{A_k}{\omega}\sin\omega t\,.
\end{eqnarray}
\end{subequations} Here we have allowed the driving amplitude $A_k$ to vary across lattice sites.

To evaluate the integrals in Eq.~\eqref{eq:H0_app}-\eqref{eq:H2_app}, it is most convenient to decompose $\calH_\text{mov}$ in terms of Fourier components
\begin{equation}
    \calH_\text{mov}(t)=\sum_m h_m e^{i m\omega t}\,,
\end{equation} where
\begin{equation}
    h_0=\hn-E_J\sum_k\frac{F_{0,k}e^{-i\hf_k}+F_{-0,k}^* e^{i\hf_k}}{2}\,,
\end{equation}
\begin{equation}
    h_{m\neq 0}= -E_J\sum_k\frac{F_{m,k}e^{-i\hf_k}+F_{-m,k}^* e^{i\hf_k}}{2}\,,
\end{equation} and $F_{m,k}$ is the Fourier coefficient of $e^{i\theta_k(t)}$:
\begin{equation}
    e^{i\theta_k(t)}=\sum_m e^{i m\omega t} F_{m,k}\,.
\end{equation} The explicit expression for $F_{m,k}$ is given in Eq.~\eqref{eq:Fmk} in the main text.

We proceed to compute the terms up to second order in the Floquet Hamiltonian. The zeroth-order result is simply
\begin{equation}
    \calH_\text{eff}^0 = h_0\,.
\end{equation}
The first-order Floquet Hamiltonian is
\begin{equation}
    \calH_\text{eff}^1 = \sum_{m,n} I^{(1)}_{mn}[h_m,h_n]=\sum_{m\neq 0}\left(I^{(1)}_{m0}-I^{(1)}_{0m}\right)[h_m,h_0]\,,
\end{equation} where in the second equality we have utilized that $[h_m,h_n]=0$ for $m,n\neq 0$. The coefficient $I_{mn}^{(1)}$ is given by the integral
\begin{equation}
    I_{mn}^{(1)}=\frac{1}{2iT} \int_0^T \rd t_1 \int_0^{t_1}\rd t_2 \exp(i\omega(m t_1+n t_2))\,,
\end{equation} and we obtain $I^{(1)}_{m0}=-I^{(1)}_{0m}=-1/(2m\omega)$. The last piece is to compute the commutator $[h_m,h_0]$, which can be readily done using the identity
\begin{align}
    [e^{i\hf_k},G(\hat{n}_k)]&=
 -e^{i\hf_k}\Delta_k G(\hat{n}_k)\,, \\
    [e^{-i\hf_k},G(\hat{n}_k)] &= \Delta_k G(\hat{n}_k) e^{-i\hf_k}\,,
\end{align}
where $G$ is an arbitrary function of $\hat{n}_k$. Here $\Delta_k G(\hat{n}_k)=G(\hat{n}_k+1)-G(\hat{n}_k)$ is the forward difference. We therefore arrive at the result
\begin{equation}\label{}
      \calH_\text{eff}^1=\frac{E_J}{2\omega}\sum_k\left[B_k\Delta_k \hn e^{-i\hf_k}+\tn{h.c.}\right]\,,
\end{equation} where
\begin{equation}
    B_k=\sum_{m \neq 0} \frac{F_{m,k}}{m}\,.
\end{equation}

Continuing on to the second order correction, the expression for the Floquet Hamiltonian is
\begin{equation}\label{}
  \calH^{2}_\text{eff}=\sum_{mnk} (I_{mnk}^{(2)}+I_{knm}^{(2)})[h_m,[h_n,h_k]]\,,
\end{equation}  where
\begin{equation}\label{}
  I_{mnk}^{(2)}=\frac{-1}{6T}\int_0^T \rd t_1 \int_0^{t_1}\rd t_2 \int_0^{t_2}\rd t_3 e^{i\Omega(mt_1+nt_2+kt_3)}\,.
\end{equation} To have a nonzero commutator, we need at least $n=0$ or $k=0$,
\begin{equation}\label{}
  \calH^{2}_\text{eff}=\sum_{mn}[h_m,[h_n,h_0]]\left(I^{(2)}_{mn0}+I^{(2)}_{0nm}-I^{(2)}_{m0n}-I^{(2)}_{n0m}\right)\,.
\end{equation}
Therefore $n\neq 0$, and the rest can be decomposed into three parts:
\begin{subequations}
\beq\label{eq:H2A_app}
  H^{2}_A&=&\sum_{m\neq 0,n\neq 0 }[h_m,[h_n,h_0]]\left(I^{(2)}_{mn0}+I^{(2)}_{0nm}-I^{(2)}_{m0n}-I^{(2)}_{n0m}\right)\,,\nn\\ \\
\label{}
  H^{2}_B&=&\sum_{n\neq 0 }[h_0-\hn,[h_n,h_0]]\left(2I^{(2)}_{0n0}-I^{(2)}_{00n}-I^{(2)}_{n00}\right)\,,\\
\label{}
  H^{2}_C&=&\sum_{n\neq 0 }[\hn,[h_n,h_0]]\left(2I^{(2)}_{0n0}-I^{(2)}_{00n}-I^{(2)}_{n00}\right)\,.
\eeq
\end{subequations}
The coefficients in the parenthesises above can be calculated and we obtain
\begin{subequations}
\beq\label{eq:IS}
    \left(I^{(2)}_{mn0}+I^{(2)}_{0nm}-I^{(2)}_{m0n}-I^{(2)}_{n0m}\right)_S &=& \frac{1}{2mn\omega^2}-\frac{1}{2m^2\omega^2}\delta_{m,-n},\quad (m,n\neq 0)\,, \nn\\ \\
    \left(2I^{(2)}_{0n0}-I^{(2)}_{00n}-I^{(2)}_{n00}\right) &=& \frac{1}{n^2\omega^2}\,,\quad (n\neq 0)\,.\nn\\
\eeq
\end{subequations}
In Eq.~\eqref{eq:IS}, we have symmetrized the result with respect to $m\leftrightarrow n$ (indicated by the subscript $S$), because  $m,n$ can be exchanged in Eq.~\eqref{eq:H2A_app} due to $[h_m,h_n]=0$. The final step is to compute the commutators, and we obtain Eqs.~\eqref{eq:H2_main} in the main text.

\section{{  Discussion of  Secondary Freezing Peaks}}
\label{App:D}

{Here we comment on some additional details associated with the freezing phenomenology. We seek to explain both the emergent conservation of $\hat{N}$ at discrete points away from freezing and how the choice of initial state affects the robustness of freezing.}

Consider for simplicity starting with ${\cal{H}}_D=2$ (i.e. equivalent to a spin$-1/2$ system). Recall that for the square-wave drive within the zeroth-order Magnus expansion, the non-integrable portion of the Hamiltonian is:
\begin{equation}\label{condition_supp1}
    \dfrac{\omega E_{J}}{A\pi}\sin \Big(\dfrac{\pi A}{\omega} \Big) \sum_{k}\cos{\hat{\varphi}_{k}}- \dfrac{\omega E_{J} }{A\pi}
   \Big(1- \cos \Big(\dfrac{\pi A}{\omega}\Big) \Big )\sum_{k}\sin{\hat{\varphi}_{k}}.
\end{equation}
The matrix elements in the basis states{  (Fock states)} of interest are
\begin{equation}
    \langle m_{k}| \cos{\hat{\varphi}_{k}}|n_{k} \rangle, \   \langle m_{k}| \sin{\hat{\varphi}_{k}}|n_{k} \rangle,
\end{equation}
Performing a Taylor expansion of the $\cos\varphi,~\sin\varphi$ terms in the transmon regime, we note that for the given ${\cal{H}}_D$, the former only contributes for $m=n$, while the latter contributes even when $m\neq n$.

If we recall the condition for freezing under the square-wave drive, we find that $A/\omega = 2\mathbb{Z}$, where both terms in Eq.~\eqref{condition_supp1} vanish. However,  in Fig.~\eqref{fig:Hilbertdepen}(a), we notice additional peaks not only at $A/\omega = 2 \mathbb{Z}$ (even integer multiples) but also at odd integer values.
For odd integer ratios of $A/\omega$, the first term in Eq.~\eqref{condition_supp1} goes to zero. Although the second term remains non-zero, its impact is less pronounced because it involves off-diagonal elements.
In Fig.~\ref{fig:Hilbertdepen}(b), we observe that as the ratio of amplitude to frequency increases, ${\cal{F}}$ approaches $1$ regardless of whether we are near or away from freezing. This is due to the effect of the nonintegrable part of the effective Hamiltonian being diminished quantitatively. It is noteworthy that a similar trend is evident in the plot of the Lyapunov exponent shown in Fig.~\ref{fig:classicalLE}(a). In Fig.~\ref{fig:Hilbertdepen}(c) we show the initial state dependence on ${\cal{F}}$ as a function of $A/ \omega$, which demonstrates  that the robustness associated with the freezing peaks for the higher excited states (see Fig.~\ref{fig:nexptH(0)}) diminishes (Eq.~\eqref{condition_supp1}). We only observe the remaining freezing peaks for ${\cal{F}}$ at the even integer ratios of $A/ \omega$.

{  \section{Dependence of Semi-Classical Freezing on $\omega$}
\label{App:semiclassical}}

Here we comment on the threshold on $\omega$ below which semi-classical freezing vanishes. In Fig. \ref{fig:omega_threshold_vs_alpha}(a) we plot the MLE for a one-dimensional chain of length $8$ due to the square-wave drive, as a function of the ratio between drive frequency $\omega$ and the transmon frequency $\nu$. The curves show the first six freezing points (as determined by the zeroth-order Magnus expansion) given by Eq.~\eqref{condn1}. We find that the $\omega$-threshold has a clear dependence on the value of $\alpha=A/\omega$ indicating that the $\omega$-threshold is itself a function of the drive amplitude $A$. Interestingly, we observe that the $\omega$-threshold seems to converge towards an $O(\nu)$ number for increasingly large values of $\alpha$.

In Fig.~\ref{fig:omega_threshold_vs_alpha}(b) we plot the MLE for the same chain and drive, as a function of the drive amplitude $A$ at fixed frequencies $\omega/(2\pi)=15$ GHz and $\omega/(2\pi) =9.32$ GHz. We observe that the first freezing point at $\alpha=2$ that is present in the former case, does not appear for the smaller drive frequency. The remaining freezing points are present for both frequencies. This is in alignment with Fig.~\ref{fig:omega_threshold_vs_alpha}(a) which indicates that the $\omega$ threshold for the first freezing point lies at around $\omega/(2\pi)\sim 11$ GHz.\\

\noindent\includegraphics[width=0.75\linewidth]{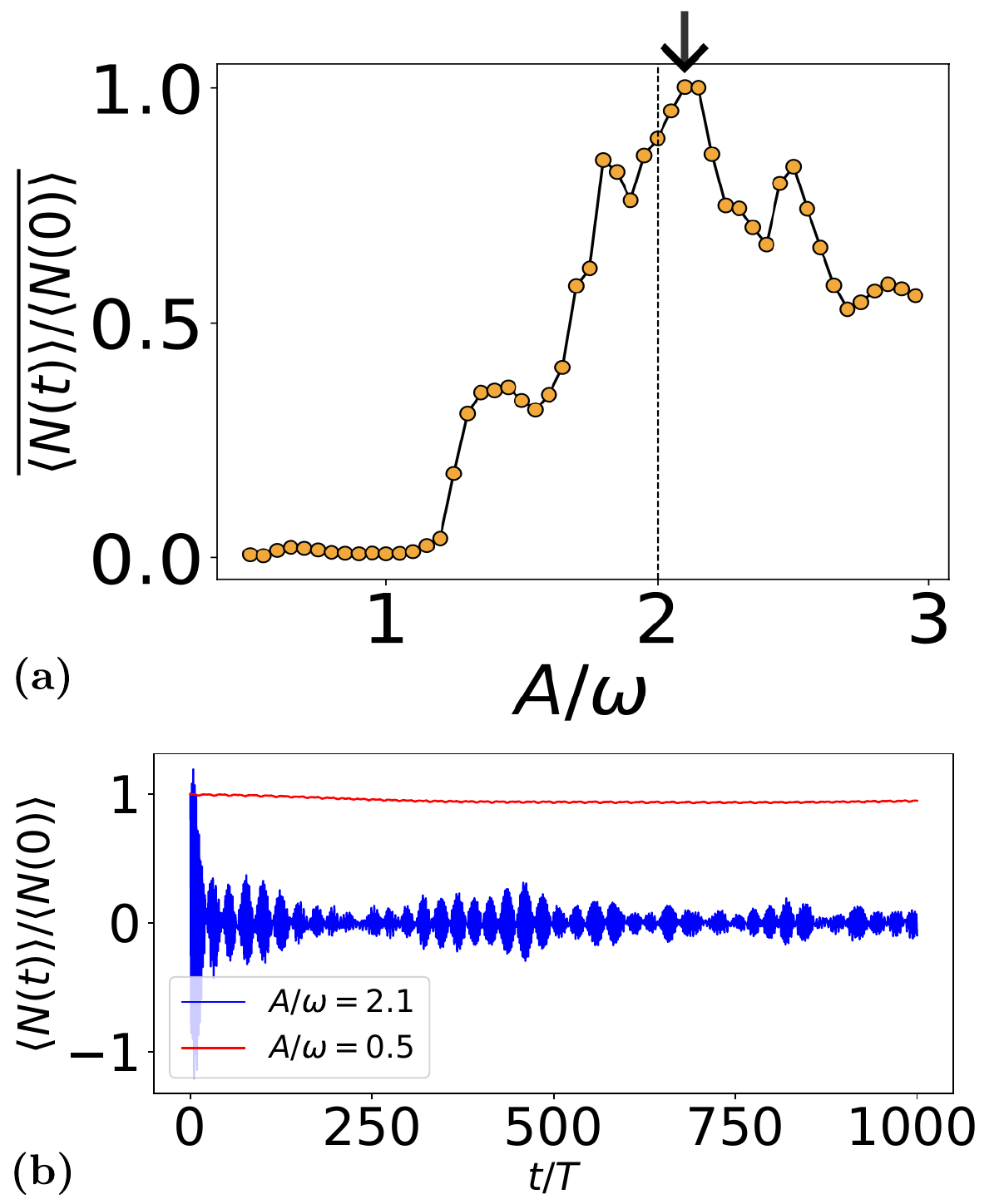}
\subsection*{Supplemental Figure 1. Freezing for large local Hilbert space dimension:}
\customlabel{fig:firstorderscaling}{S1}
(a) Time averaged expectation of $N(t)/N(0)$ is plotted as a function of $A/ \omega$ for a linear chain of $4$ transmons and local Hilbert space dimension $8$ for the square wave drive. All of the other parameters are as in Fig.~\ref{fig:ndtstrobo} in the main text. Note that there is a slight shift of the freezing point due to higher order corrections in the Magnus expansion. (b) Evolution of $\hat{N}(t)$ for the same system at and away from freezing. Note that we use the renormalized freezing point, $A/\omega=2.1$, determined numerically from (a).

\noindent\includegraphics[width=0.95\linewidth]{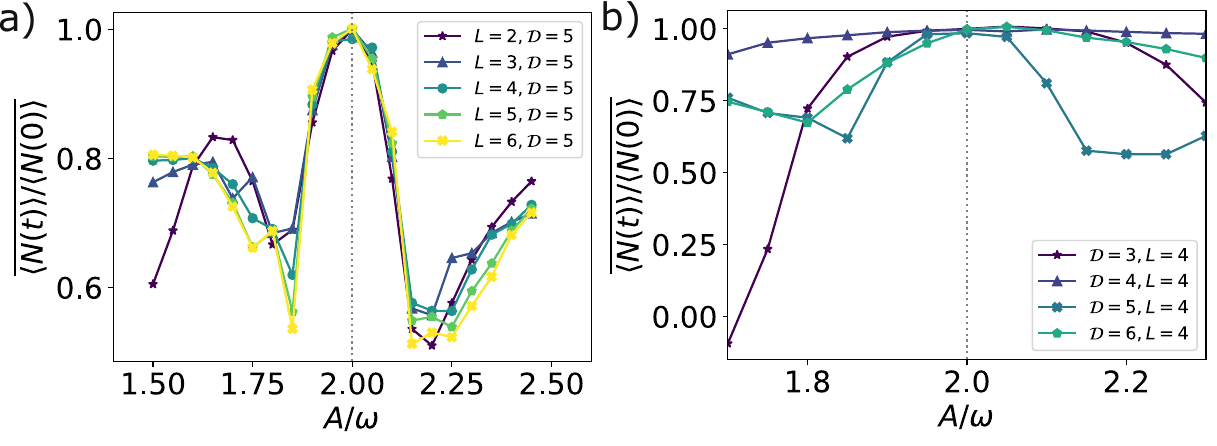}
\subsection*{Supplemental Figure 2: Dependency of freezing on local Hilbert space dimension and system size:}  
\customlabel{fig:HilbertspaceSystemsize}{S2}
(a) The time averaged (over 100 Floquet cycles) expectation of $N(t)/N(0)$ as function of $A/\omega$ for different system size ($L$) keeping the Hilbert space dimension ($\cal{D}$) fixed. (b) Same quantity for different $\cal{D}$ for fixed $L$. In both cases the $\cos(\phi)$ nonlinearity are kept up to order 8. As an initial state we chose the ground state of $H(0)$. All of the other parameters are as in Fig.~\ref{fig:ndtstrobo} in the main text.

\noindent\includegraphics[width=1.0\linewidth]{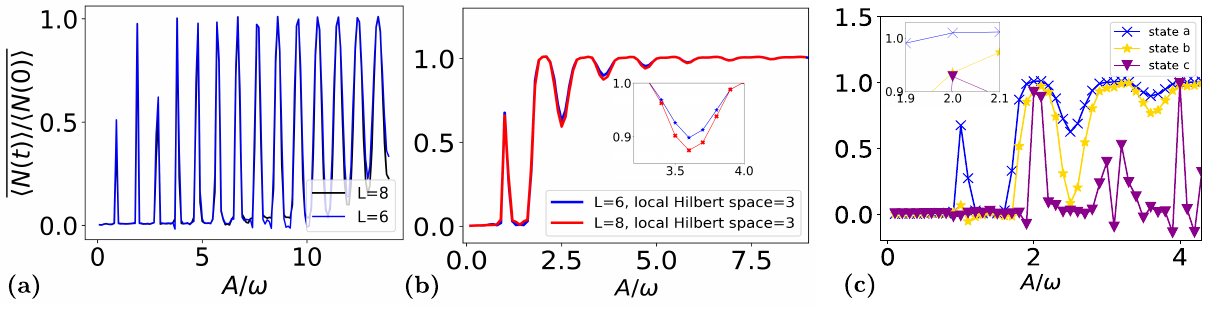}
\subsection*{Supplemental Figure 3: Behavior of secondary freezing points:}
\customlabel{fig:Hilbertdepen}{S3}
(a) ${\cal{F}}$ (Eq.~\eqref{Eq.Fidelity}) averaged over 100 Floquet cycles as a function of $\alpha=A/\omega$ for two different system sizes ($L = 6, 8$) and onsite Hilbert space (${\cal{H}}_D=2$), with $\ket{\psi(0)}$ chosen as ground state of $H(0)$. In addition to the sharp peaks at {\it even} integer $\alpha$, peaks at {\it odd} integers are also seen. (b) Same as in (a), but with ${\cal{H}}_D=3$. With increasing $\alpha$,  $\cal{F}$ flattens out since $\h^0_{\widehat{\varphi}}$ weakens compared to $\h^0_{\widehat{n}} + f(t)\h^{\rm{drive}}_{\widehat{n}}$. Furthermore, as anticipate, we see a reduction in the extent of the conservation of $\hat{N}$ away from freezing for the larger system size. (c) For $L=8$ transmons and ${\cal{H}}_D=3$, variation of ${\cal{F}}$ for different $\ket{\psi(0)}$, drawn from the eigenspectrum of $H(0)$ in Fig.~\ref{fig:nexptH(0)}.
All other parameters are as in Fig.~\ref{fig:ndtstrobo} in the main text.

\noindent\includegraphics[width=0.75\linewidth]{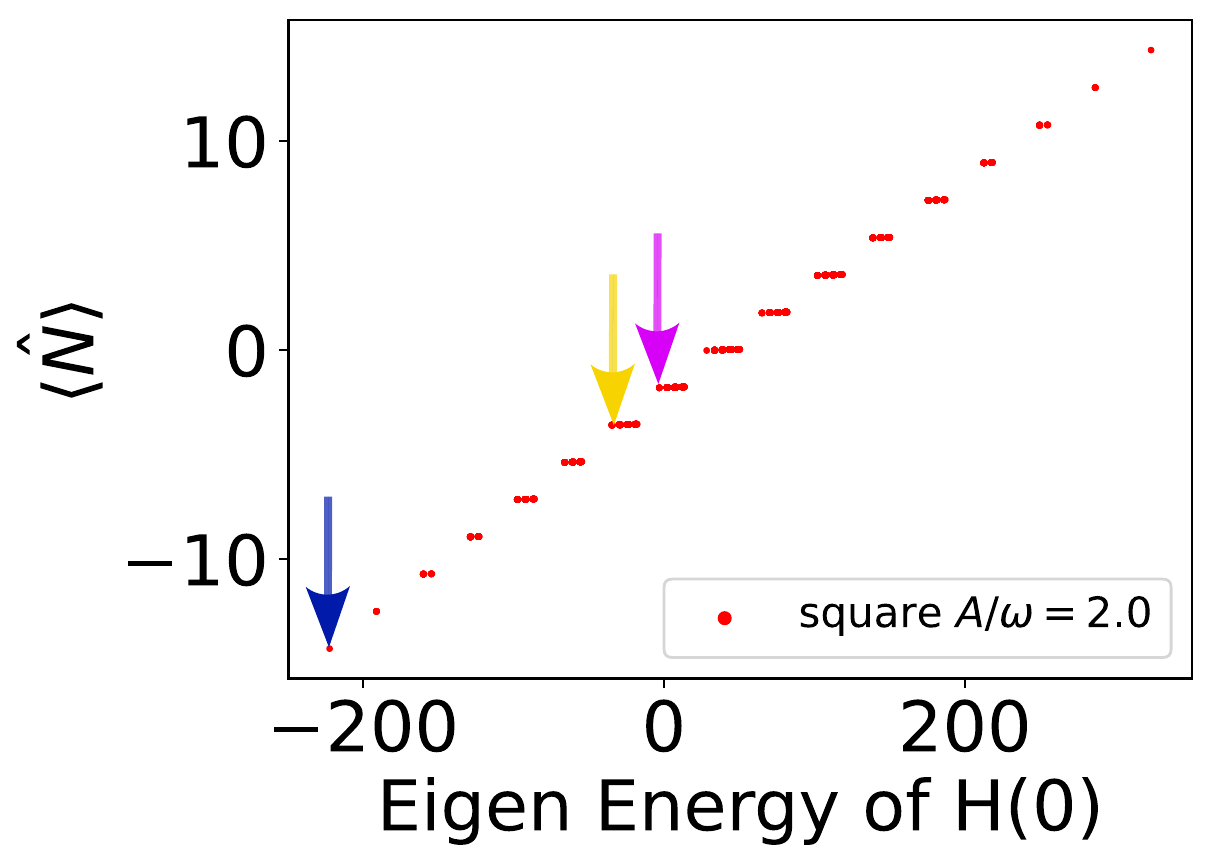}
\subsection*{Supplemental Figure 4: $\hat{\textbf{N}}$ expectation of initial states}
\customlabel{fig:nexptH(0)}{S4}
The expectation value of the operator $\hat{N}$ is calculated with respect to the eigenstates of the Hamiltonian $H(0)$ and plotted against the corresponding eigen-energy for square wave drive.} The eigenstates marked by arrows are used as $\ket{\psi(0)}$ to calculate the observables in Eq.~\eqref{Eq.Fidelity} and Fig.~\ref{fig:Hilbertdepen}.

\noindent\includegraphics[width=0.95\linewidth]{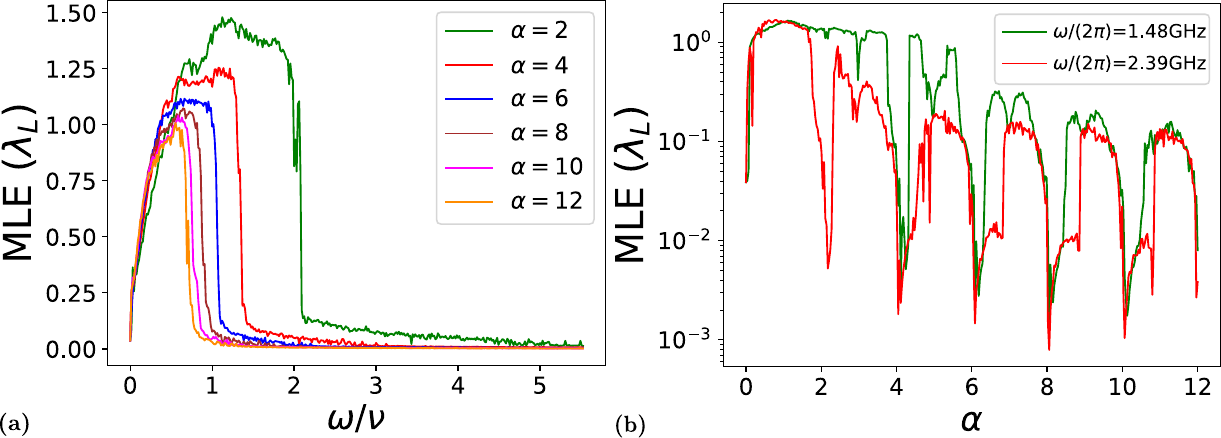}
\subsection*{Supplemental Figure 5: Frequency threshold of semi-classical freezing:}
\customlabel{fig:omega_threshold_vs_alpha}{S5}
(a) MLE plotted as a function of $\omega/\nu$ at various zeroth-order freezing points of the square-wave drive. The system is a length $8$ chain initialized as $1010\ldots$. For larger freezing points, $\alpha=A/\omega$, the omega threshold, below which freezing does not occur, decreases while appearing to approach an $O(\nu)$ number. (b) MLE plotted as a function of $\alpha$ at fixed frequencies $\omega/(2\pi)=15$ GHz and $\omega/(2\pi) =9.32$ GHz of the square-wave drive. Notably, the $\alpha=2$ freezing point is absent in the latter case indicating that the omega threshold for the freezing point lies between $9.32$ and $15$ GHz.

\bibliography{Reference.bib}

\end{document}